\newif\ifarxiv
\arxivtrue

\ifarxiv
\documentclass{jfm-arxiv}
\else
\documentclass[lineno]{jfm}
\fi

\usepackage{graphicx}
\usepackage{epstopdf,epsfig}
\usepackage{newtxtext}
\usepackage{newtxmath}
\usepackage{natbib}
\usepackage{hyperref}
\usepackage{amsmath}
\usepackage{caption}
\usepackage{subcaption}
\usepackage{float}
\usepackage{multicol}

\usepackage{array}
\newcommand{\PreserveBackslash}[1]{\let\temp=\\#1\let\\=\temp}
\newcolumntype{C}[1]{>{\PreserveBackslash\centering}p{#1}}

\usepackage[normalem]{ulem}
\hypersetup{
    colorlinks = true,
    urlcolor   = blue,
    citecolor  = black,
}

\newcommand{\RomanNumeralCaps}[1]
\linenumbers

\title{The stability of stratified horizontal flows of carbon dioxide at supercritical pressures}

\author{Marko Draskic\aff{1},
  Jerry Westerweel\aff{1}
 \and Rene Pecnik\aff{1}\corresp{\email{r.pecnik@tudelft.nl}}}

\affiliation{\aff{1}Department of Process $\&$ Energy - ME, Delft University of Technology, Leeghwaterstraat 39, 2628 CB Delft, The Netherlands}

\begin{document}
\maketitle

\begin{abstract}
Fluids at supercritical pressures exhibit large variations in density near the pseudo critical line, such that buoyancy plays a crucial role in their fluid dynamics. Here, we experimentally investigate heat transfer and turbulence in horizontal  hydrodynamically developed channel flows of carbon dioxide at $88.5$ bar and $32.6^{\circ}C$, heated at either the top or bottom surface to induce a strong vertical density gradient. In order to visualise the flow and evaluate its heat transfer, shadowgraphy is used concurrently with surface temperature measurements. With moderate heating, the flow is found to strongly stratify for both heating configurations, with bulk Richardson numbers $Ri$ reaching  up to 100. When the carbon dioxide is heated from the bottom upwards, the resulting unstably stratified flow is found to be dominated by the increasingly prevalent secondary motion of thermal plumes, enhancing vertical mixing and progressively improving heat transfer compared to a neutrally buoyant setting. Conversely, stable stratification, induced by heating from the top, suppresses the vertical motion leading to deteriorated heat transfer that becomes invariant to the Reynolds number. The optical results provide novel insights into the complex dynamics of the directionally dependent heat transfer in the near-pseudo-critical region. These insights contribute to the reliable design of heat exchangers with highly property-variant fluids, which are critical for the decarbonisation of power and industrial heat. However, the results also highlight the need for further progress in the development of experimental techniques to generate reliable reference data for a broader range of non-ideal supercritical conditions. 
\end{abstract}


\section{Introduction}

\begin{figure}
    \centering
    \begin{subfigure}[t]{0.336\textwidth}
         \centering
         \includegraphics[width=\textwidth]{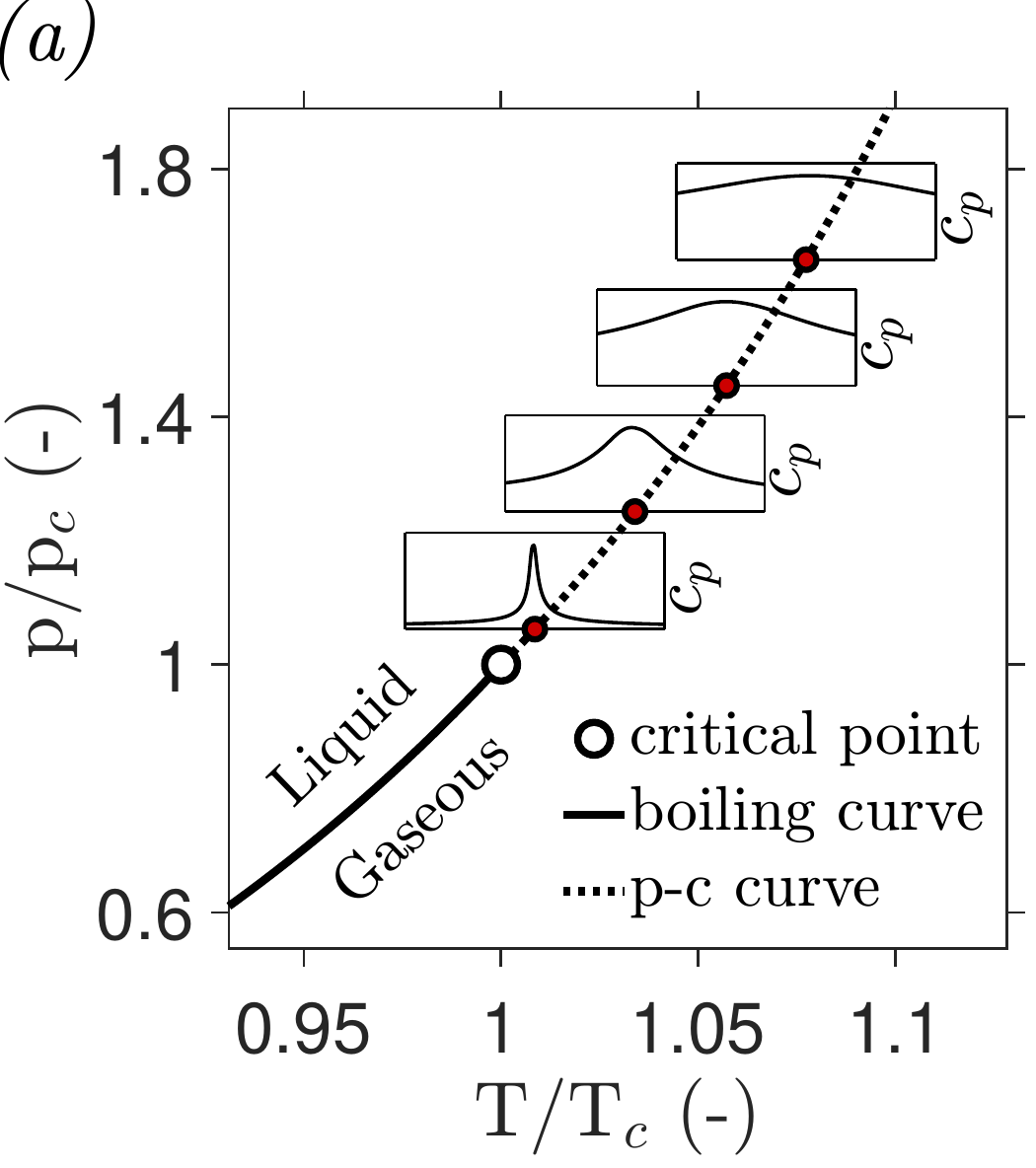}
         %\label{fig: prop1}
    \end{subfigure}
    \hfill
    \begin{subfigure}[t]{0.32\textwidth}
         \centering
         \includegraphics[width=\textwidth]{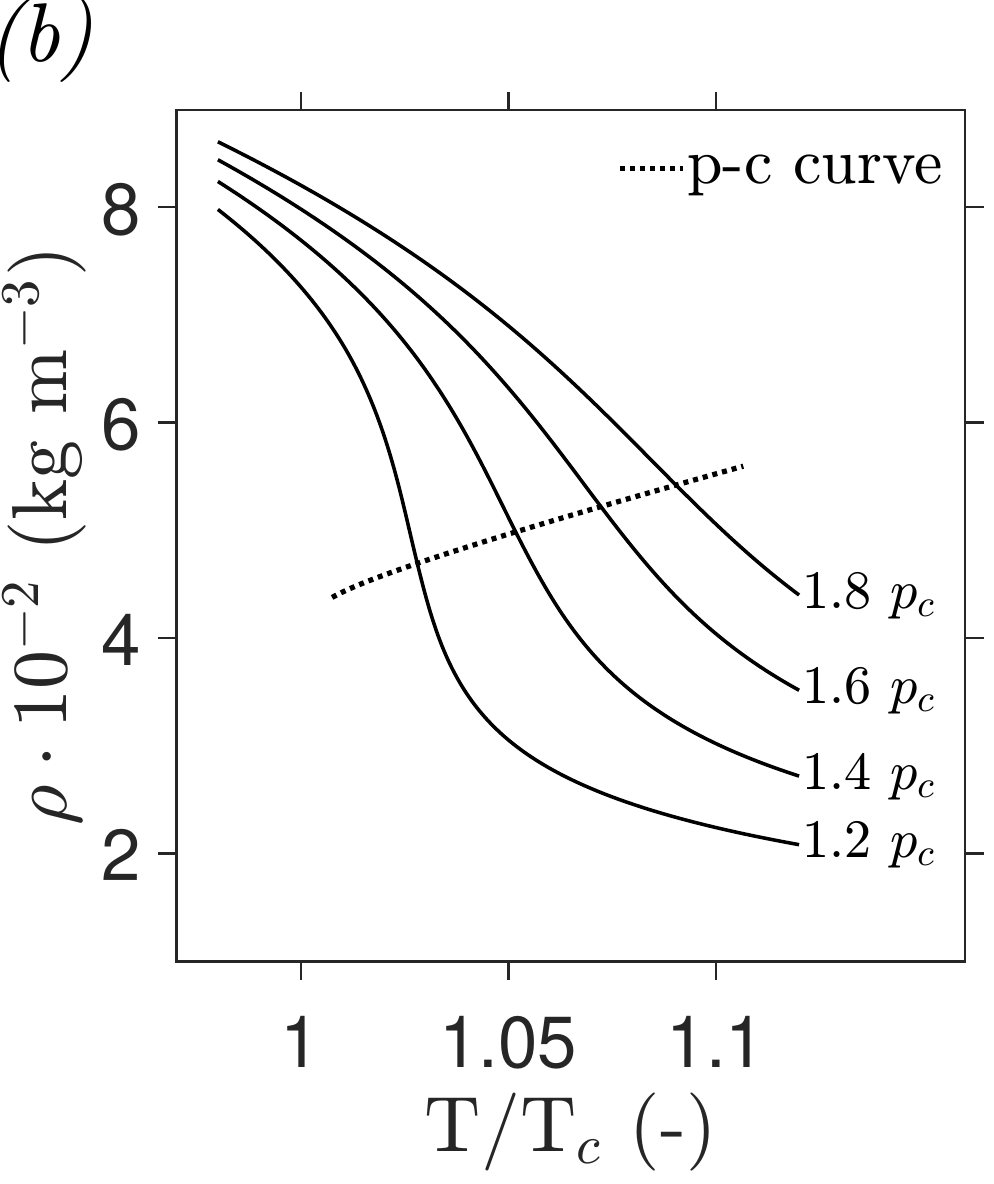}
         %\label{fig: prop2}
    \end{subfigure}
    \hfill    
    \begin{subfigure}[t]{0.32\textwidth}
         \centering
         \includegraphics[width=\textwidth]{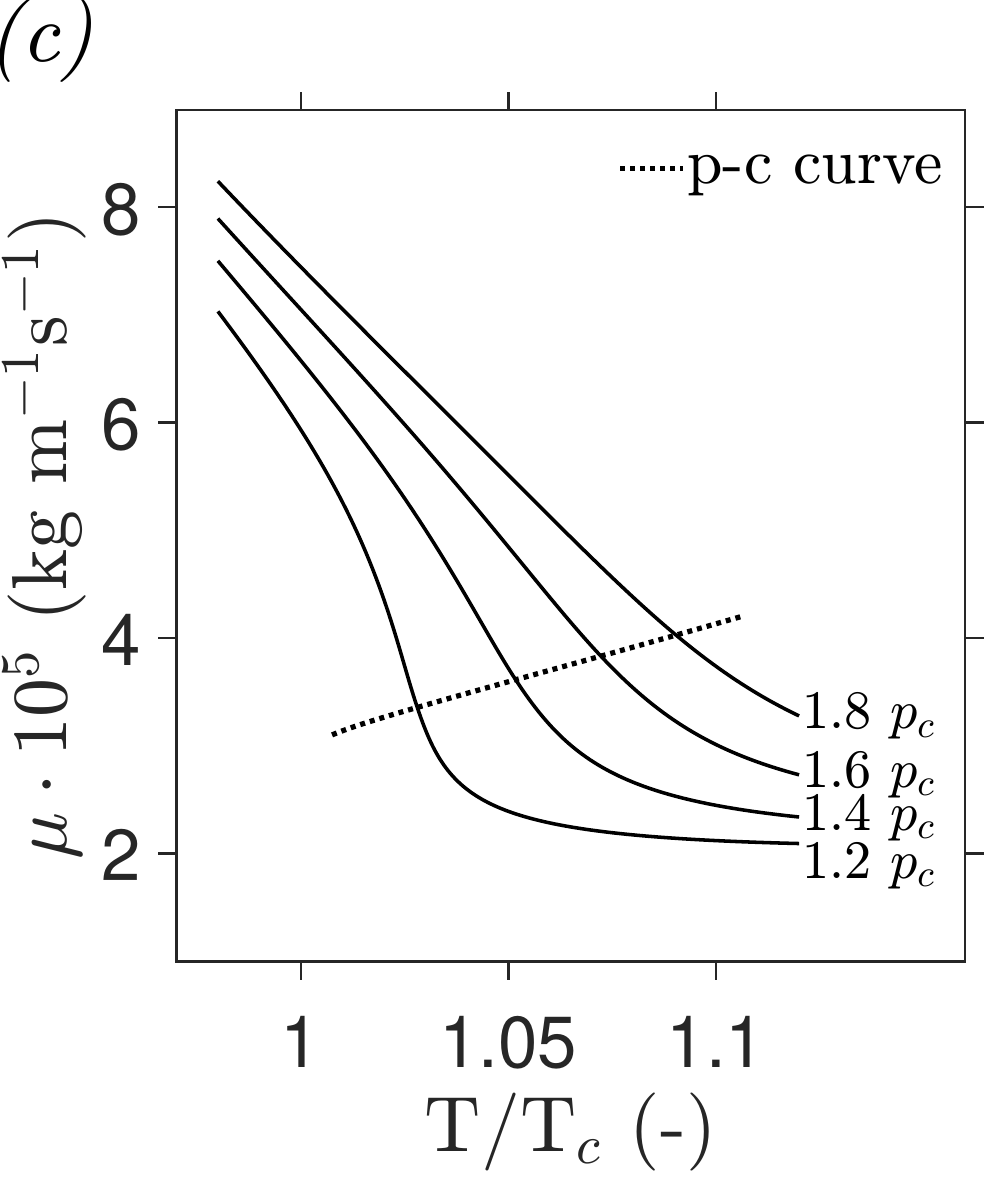}
         %\label{fig: prop3}
    \end{subfigure}
\caption{Thermodynamic properties of carbon dioxide at supercritical pressures, for which the critical temperature and pressure are at $T_{\text{c}}=31.0^{\circ}C$ and $p_{\text{c}}=73.8$ bar, respectively. Subfigure (a) shows normalised profiles of the isobaric heat capacity $c_{\text{p}}$ along several isobars, the pressure of which is indicated in red. The pseudo-critical (p-c) curve is defined at the local maxima of $c_{\text{p}}$ along an isobar \citep{banuti2015crossing}. Distributions of density $\rho$ and viscosity $\mu$ near the p-c line are shown for several supercritical pressures in (b) and (c), respectively.}
\label{fig:thermodynamic properties}
\end{figure}
Beyond the vapour-liquid critical point, a fluid no longer undergoes a discrete phase transition from liquid to gas when it is heated. Instead, at supercritical pressure, fluids undergo a boiling-like process in which they remain in a continuous phase. During this pseudo-boiling process, fluids exhibit considerable non-linear variations of thermodynamic properties. The thermodynamic gradients are particularly large near the pseudo-critical temperature $T_{\text{pc}}$, defined at the temperature where the specific isobaric heat capacity has its maximum along a supercritical isobar \citep{banuti2015crossing}. The location of the pseudo-critical (p-c) line that spans all pseudo-critical temperatures is indicated in figure~\ref{fig:thermodynamic properties}. Several novel energy conversion systems are currently being designed to operate with fluids at supercritical pressures, as they can yield higher conversion efficiencies and compacter designs \citep{brunner2010applications}. Therewith, supercritical energy systems are promising solutions for the generation of power from non-fossil heat sources, for compact district heating systems, and for the generation of high-temperature industrial heat. However, as a result of the large thermophysical variations in these systems, the flows within them differ considerably from those in their counterparts with sub-critical single-phase fluids. 

At a supercritical pressure, the fluid dynamics of heated or cooled fluids are often strongly affected by buoyancy. In this complex thermodynamic region, moderate heating rates can already induce significant variations in density that lead to sharp, local flow accelerations or mixed convection when buoyancy becomes significant. Therefore, heat transfer in engineering application, which operate with supercritical fluids, often depends on the direction of gravity. 
% Therefore, the near-pseudo-critical heat exchangers of supercritical energy systems are highly susceptible to thermal stratification, and their heat transfer is often directionally dependent as a result. 
These effects can be significant, as noted in the review on heat transfer at supercritical pressure by \citet{wang2023critical}. In their work, a comparison of the empirical heat transfer rates of various experimental facilities showed that heat transfer rates can either increase or decrease (depending on the direction of the heat transfer) by up to an order of magnitude with respect to a neutrally buoyant setting. This effect is most notably prevalent in horizontal flows with vertical gradients of density, in which turbulence can be either enhanced or suppressed by buoyancy. In such settings, in particular with fluids in proximity of the pseudo-critical region, it is essential to understand how buoyancy modulates the flow and turbulence to accurately capture and predict trends in heat transfer. 

So far, mixed convection has been studied extensively for fluids that follow the Oberbeck-Boussinesq (OB) approximation. Contrary to a fluid at a supercritical pressure, the density of OB fluids is assumed to vary only moderately and linearly with temperature. Therefore, large temperature differences or large length scales are required to induce strong stratifications in the conventional single-phase fluids that are accurately described with the OB approximation \citep{zonta2018}.

The unstable stratification of a horizontal flow of an OB fluid, in which a lighter layer is formed below a denser bulk flow, is characterized by ejections of warmer fluid away from the heated wall, and the concurrent sweeping of colder fluid in the opposite direction \citep{garai2014}. However, unlike in purely convective flows, the ejected plumes form roll vortices towards the bulk of the channel in the direction of the forced convection \citep{garai2014,scagliarini2014heat,pirozzoli2017}. In the direct numerical simulations (DNS) of \citet{garai2014}, the wide roll vortices are seen to align after one another in the streamwise direction. Furthermore, the DNS of both \cite{dongli2017} and \cite{zonta2014} have shown that the intensity of the near-wall turbulence and the mixing efficiency increase when roller ejection and sweep processes are incited in a previously neutrally buoyant channel flow. Here, the improvements in mixing efficiency are diminished as the relative strength of the forced convection is increased \citep{zonta2014}. Moreover, the concurrent processes in which rollers are ejected from and swept towards the bottom surface become the dominant mechanism of wall-normal heat transfer in mixed convective flows \citep{garai2014,blass2020flow}. As such, the wall-fluid heat transfer of flows with buoyant plumes differs from - and far exceeds - the heat transfer of neutrally buoyant forced convection. 

Distinctly different behaviour is observed and simulated in stably stratified flows of OB fluids, in which a lower density layer is formed on top of a denser bulk flow. Here, fluid parcels of a certain density are preferentially redistributed along the vertical density gradient that is imposed on the flow \citep{lienhard1990}. If the gradient is sharp enough, the potential energy required to stir fluid parcels away from their preferential vertical positions is not met \citep{garcia-villalba2011}. As a result, overturns and vertical thermal structures are elongated and eventually aligned with the direction of the flow \citep{williams2017,smith2021,zonta2022}. The resulting restriction of vertical momentum transfer has been noted to greatly reduce turbulent mixing and near-wall heat transfer \citep{ohya1997,garcia-villalba2011,williams2017}, in both numerical studies and experiments. Often, the distinction is made between weak and strong stratification, following the findings of \citet{gage1968}. In the weak limit, turbulence is sustained close to the wall where shear overcomes buoyancy \citep{zonta2022}, as the latter preferentially acts on larger scales of vertical motion \citep{lienhard1990,zonta2018}. In the case of strong stratification, a global turbulent state cannot be sustained \citep{nieuwstadt2005,zonta2018}. The flow is said to become intermittent, with regions of complete turbulence suppression. For higher values of the Reynolds number $\Rey$ the formed laminar patches are more confined to the top wall, instead of spanning the entire domain, according to the results of \citet{deusebio2015}.

Unfortunately, the amount of works that consider the stratification of fluids with thermodynamic gradients beyond the limits of the Oberbeck-Boussinesq approximation has so far remained limited. However, it is with such fluids, most notably so for fluids at a supercritical pressure, that strong stratifications most readily prevail in practical applications. The influence of buoyancy on the flow of a fluid at a supercritical pressure have been studied in the context of vertical pipe flows \citep{bae2005direct, nemati2015mean} and for vertical annular flows \citep{peeters2016}. Additionally, \citet{valori2019} have optically explored a purely convective flow of supercritical carbon dioxide. The effect of buoyancy on a horizontal supercritical flow has first been considered by \citet{chu2016flow}. In their work, \citet{chu2016flow} have numerically (DNS) investigated the non-axisymmetric temperature profiles in heated horizontal pipe flows previously found in experimental investigations of pipe surface temperatures \citep{adebiyi1976experimental,yu2013experimental,tian2021staged,cheng2024supercritical}. Here, the overall deterioration in heat transfer with respect to a neutrally buoyant flow \citep{theologou2022experimental,cheng2024supercritical} was attributed to the accumulation of a light, warmer layer at the top of the horizontal pipe \citep{chu2016flow}. Furthermore, the thermal stratification of the flow of a supercritical fluid through a heat exchanger channel has been touched upon in the DNS of \citet{wang2023direct}. Here, the initially turbulent flow was predicted to laminarize along the imposed density gradient, subsequently deteriorating its wall-normal heat transfer. A first experimental investigation of horizontal buoyancy-affected flows of supercritical $\text{CO}_2$ was presented by \citet{whitaker2024flow}. Here, side-view schlieren imaging was used to study the flow within a heated microchannel. In their work, the optical signal was greatly affected by the direction of the heat transfer. Nevertheless, only a moderate difference between the the heat transfer of the bottom-up heated surface and the top-down heated surface were measured for their operating conditions. Therewith, the effects of strong stratifications on the heat transfer of continuous, mixed convective flows at supercritical pressures remain experimentally unexplored.

The current study experimentally investigates horizontal flows of carbon dioxide at supercritical pressure in which buoyancy is non-negligible, or even becomes dominant. More specifically, initially hydrodynamically developed steady horizontal channel flows are considered, subject to one-sided heating such that vertical gradients of density are induced. The large, non-ideal variations in thermodynamic properties of the heated carbon dioxide make that the Oberbeck-Boussinesq approximation does not apply, and that significant buoyant contributions can be induced at the moderate heating rates that are attainable in laboratory scale experiments. As such, the modulation of turbulence in both stably and unstably stratified channel flows can be addressed in the current work. 

To visualize the  stratifications of carbon dioxide at supercritical pressure, the surface temperature measurements are complemented with optical diagnostics in this work. The optical tools exploit variations in refractive index in the working fluid to yield visualisations of compressible turbulent carbon dioxide. Contrary to the investigations of wall temperature, the resolution of these optical measurements is not limited in time by the thermal inertia of the system, or in space by the finite physical size of probes and probe wells. Therewith, the current optical methods can partially fill the knowledge gap left by the incompatibility of high-frequency hot-wires with non-ideal media \citep{vukoslavvcevic2005}. As such, the qualitative visualisations are used to explain trends in the wall temperature data by revealing highly intermittent local flow patterns previously unexplored in experiments with fluids at supercritical pressures.

A description of the experimental facility and the (optical) measurement tools is given in \S \ref{sec:Methodology}. In \S \ref{sec:results}, the applicability of the optical diagnostics is first evaluated by considering the unheated base flow. Thereafter, visualisations of heated channel flows in both stably and unstably stratified configurations are presented and discussed. Furthermore, surface temperature data are presented for both configurations. Finally, a summary of the most important conclusions is presented in \S \ref{sec:conclusions}. At the end of this document, supplementary information on the used optical filters is given in appendix \ref{sec:appA}.

\section{Methodology}\label{sec:Methodology}
\begin{figure}
    \centering
\includegraphics[width=1.025\textwidth]{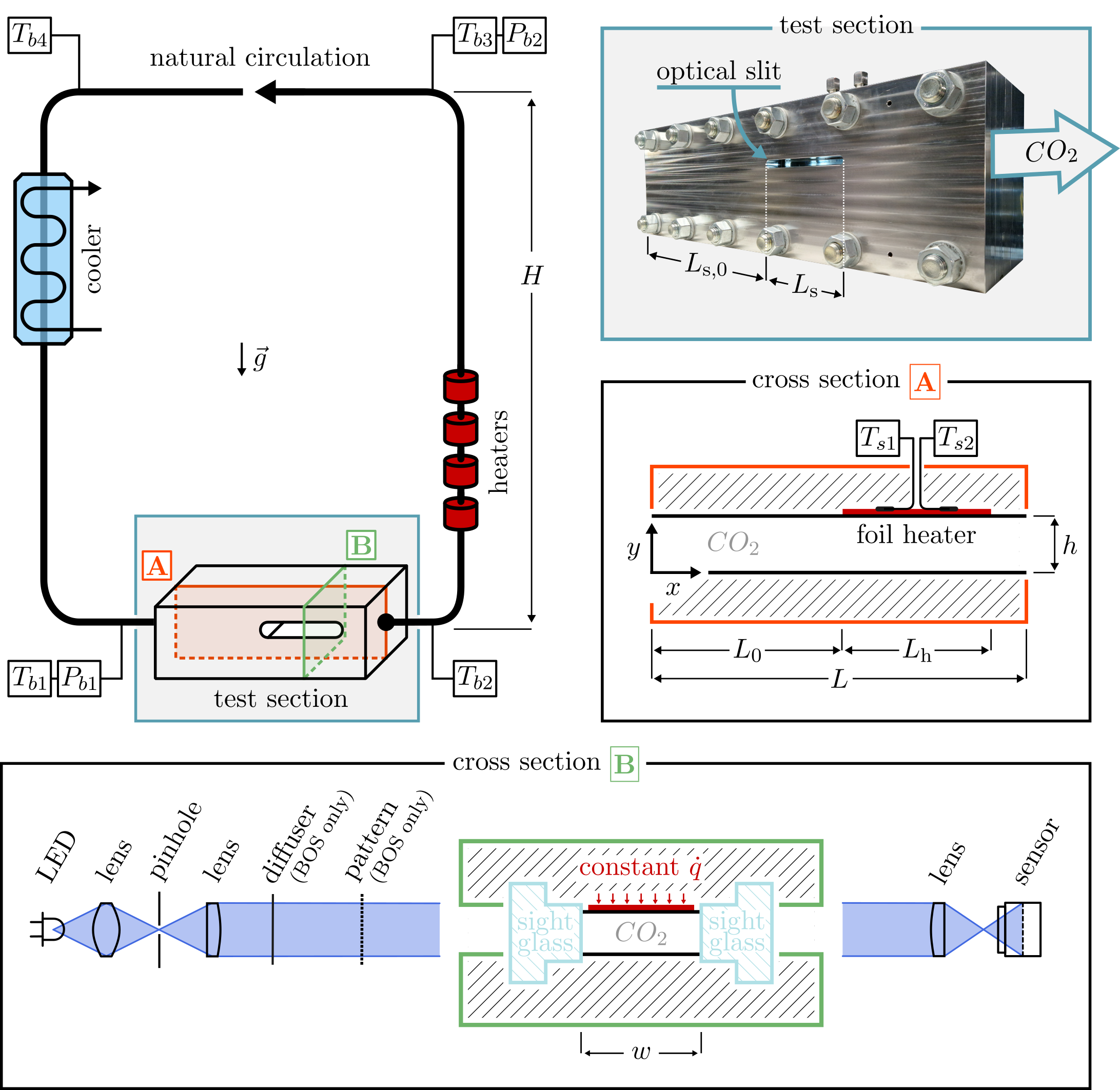}
\caption{Schematic of the current experimental facility, in which carbon dioxide at supercritical pressures is naturally circulated to a test section that is bi-laterally optically accessible. An image of the test section is outlined in blue. In the image, any usually present thermal insulation (30 mm polyethylene foam) has been removed. A schematic of lateral cross section~\textbf{A} is outlined in orange. In order to change the direction of the heating with respect to the schematic, the test section is inverted. The current optical configuration is shown alongside green-outlined cross section~\textbf{B}. Any symbols in the figure are explained in the main text, and in table \ref{tab:setup description}. }
\label{fig:schematic}
\end{figure}
\subsection{Experimental facility}\label{sec:meth:facility}
In this work, the stratification of a non-ideal fluid is studied experimentally in a novel test facility that provides optical access to a continuous flow of carbon dioxide at supercritical pressures. The experimental facility is shown schematically in figure~\ref{fig:schematic} and an overview of the geometric parameters is given in table~\ref{tab:setup description}. 
\subsubsection{Natural circulation}
A steady, continuous flow of carbon dioxide is generated using natural circulation. The current Natural Circulation Loop (NCL) is shown schematically in the top-left corner of figure~\ref{fig:schematic}. In the NCL, the non-linear gradients in density of the $\text{CO}_2$ are exploited to generate a differential buoyant force over the vertical legs of the NCL, therewith inducing a natural circulation. Contrary to most forced convective systems, NCLs do not require pumps that contain lubricants that may contaminate the working fluid and therefore limit its optical accessibility. Furthermore, the inherent induction of mechanical noise, and the intrinsic presence of leakages of mechanical propulsion are avoided. Additionally, the NCL allows for the test section pressure, its inlet bulk temperature, the imposed mass flow rate, and the foil heating rate to be individually varied. A detailed description of the NCL and its steady state performance is given by \citet{draskic2024}. 
\subsubsection{Test section}
A high-pressure stainless steel test section lies at the core of the experimental facility. As shown in the top-left of figure~\ref{fig:schematic}, the test section is part of the bottom horizontal section of the NCL. An image of the apparatus, outlined in blue, is shown in the top-right of figure~\ref{fig:schematic}. The test section provides bi-lateral optical access to a rectangular channel through tempered obround borosilicate visors, as shown in cross section~\textbf{B} in figure~\ref{fig:schematic}. Cross section~\textbf{A} shows the side-view of the rectangular channel between two gradual round-to-rectangular reducers. The reducers serve to connect the channel to the round pipe-connections of the experimental facility. The top surface of the rectangular channel is heated after a length $L_0$. In order to apply heat from the opposite, bottom surface, the test section is flipped in its entirety. However, in the flipped, bottom-up heating configuration the origin of the considered coordinate system remains on the bottom surface of the test section, consistent with cross section~\textbf{A} of figure~\ref{fig:schematic}. The heating is provided by a self-adhesive 40 W polyimide foil resistance heater that spans the full width of the channel. In order to avoid a forward facing step of the surface, the foil heater is bonded to the rear side of a removable surface plate that spans the full top surface, as indicated using the top black line in cross section~\textbf{A} of figure~\ref{fig:schematic}. The surface plate, a 0.8mm thick polished aluminium surface mirror, is oriented towards the carbon dioxide with its polished side. The rear-side surface temperature of the foil heater is measured using two K-type thermocouple probes, mounted in the lateral middle of the channel at different streamwise locations, as indicated with $T_{\text{s}1}$ and $T_{\text{s}2}$ in cross section~\textbf{A} of figure~\ref{fig:schematic}. 

\begin{table}
  \begin{center}
\def~{\hphantom{0}}
    \begin{tabular}{llC{0.3cm}rr}
        \multicolumn{2}{l}{\textbf{parameter $\&$ description}} && \textbf{value} & \textbf{unit}\vspace{0.15cm}\\
        $h$  & channel height && 7.5 & mm \\
        $w$  & channel width && 50 & mm \\
        $L$  & inter-reducer channel length && 375 & mm \\
        $L_0$ & initial unheated length && 300 & mm \\
        $L_{\text{h}}$ & heated length && 50 & mm\\
        $L_{\text{s,}0}$ & reducer-to-optical slit distance && 250 & mm\\
        $L_{\text{s}}$ & optical slit length && 140 & mm\\
        $H$ & natural circulation loop height && 4 & m\\
    \end{tabular} 
  \caption{Description of the experimental facility. The table gives the values of the parameters indicated in figure~\ref{fig:schematic}.}
  \label{tab:setup description}
  \end{center}
\end{table}
\subsubsection{Instrumentation}
To continuously monitor the facility, it is equipped with a series of sensors that are indicated throughout figure~\ref{fig:schematic}. To measure bulk temperatures throughout the natural circulation loop, Pt100 thermometers are radially inserted into the flow. These resistance thermometers, labeled $T_{\text{b}1}$ - $T_{\text{b}4}$ in figure~\ref{fig:schematic}, have a nominal accuracy of $\pm0.1^\circ C$. Absolute pressure measurements $P_{\text{b}1}$ and $P_{\text{b}2}$ are taken using welded STS ATM.1st sensors, with a nominal uncertainty of $\pm0.16$ bar or $0.1\%$. Thermocouples $T_{\text{s}1}$ and $T_{\text{s}2}$ yield measurements of the foil heater surface temperature with a nominal uncertainty of $\pm0.5^\circ C$. The thermocouples are calibrated and the readings are shifted to match the readings of resistance sensor $T_{\text{b}1}$ \textit{in situ}. The steady state mass flow rate $\dot{m}$ is deducted from the heating rate $\dot{Q}_{\text{imp}}$ imposed by the vertically placed heaters in figure~\ref{fig:schematic} and the increase in enthalpy $h$ across the heater, evaluated over the vertical leg of the NCL, i.e. 
\begin{equation}
    \dot{m}=\frac{\dot{Q}_{\text{imp}}}{h(T_{\text{b}3},P_{\text{b}2})-h(T_{\text{b}2},P_{\text{b}1})}.
    \label{eq:mdot}
\end{equation}
As found in the experiments of \citet{draskic2024}, the mass flow rate estimated by equation~(\ref{eq:mdot}) is in close agreement with the readings of a Coriolis mass flow meter when the system is at a steady state, when the pressure drop over the horizontal sections of the NCL is limited at moderate bulk velocities, and when the loop is insulated with a 40 mm layer of mineral wool to minimize heat losses. As these conditions are met in the current experiments, the Coriolis mass flow meter is removed to avoid any unnecessary disturbances to the flow, and equation~(\ref{eq:mdot}) is used to determine the mass flow rate within the NCL.

The uncertainty of all results shown in this work is evaluated by assuming the independence of the respective quantities measured by the sensors in the system. As a result, the standard deviation of any considered quantity is evaluated from the standard deviations of sensor data used to determine the considered quantity. For instance, the standard deviation $\sigma_{\text{h}}$ for $h(T,P)$ is obtained using 
\begin{equation}
    \sigma_{\text{h}}=\left[\left(\frac{\partial h}{\partial T}\right)^2\cdot\sigma_{\text{T}}^2+\left(\frac{\partial h}{\partial P}\right)^2\cdot\sigma_{\text{P}}^2\right]^{1/2}.
\end{equation} Here, $\sigma_T$ and $\sigma_P$ denote the standard deviations of process measurements $T$ and $P$, respectively.
\subsection{Optical methodology}\label{sec:meth:optics}
\subsubsection{Shadowgraphy}
A parallel-light shadowgraphy configuration is used to yield shadowgrams, as is shown schematically in cross section $\textbf{B}$ of figure~\ref{fig:schematic}. The object plane for the shadowgraphy is at the channel half-width. The brightness of the resulting shadowgram varies proportionally to the second spatial derivative of the path-integrated refractive index field \citep{settlesbook,merzkirch2012flow}. As the refractive index is proportional to the fluid density \citep{MICHELS1937995}, variations in density of the compressible carbon dioxide can be recorded. In the current setting, the shadowgrams outline the boundaries of depth-integrated thermal structures with consecutive maxima and minima in image brightness. In order to increase the sensitivity of the shadowgrams, to reduce the distortion of shadows, and to reduce geometric blur with respect to non-parallel light shadowgraphy \citep{settlesbook}, the light source (a monochromatic blue LED with a nominal wavelength of 455 nm) is collimated in the current optical setup. As imperfections in the sight glass surfaces can be perceived in these collimated-light shadowgrams, a spatial digital image filter is applied to the images shown in this work to highlight only the compressible structures of interest. The filter is further elaborated on in appendix \ref{sec:appA}.

A Pixelink PL-D755MU-T CMOS camera (2448$\times$2048 pixels, monochromatic) is used to record images. The exposure time of the sensor is maintained at 20 $\mu$s for all shadowgrams. The contrast of the recorded images is stretched to cover the full brightness range. Both the image sensor and the light collimator are traversed in the streamwise direction using synchronised linear traversing stages on either side of the test section to visualize the $\text{CO}_2$ at any position along the optical slit.

\subsubsection{Shadow image velocimetry}
A comparison of consecutive shadowgrams can reveal flow patterns within a compressible flow. When the currently studied flow is both turbulent and refractive, it is naturally seeded by eddies that travel at the local convective speed of the flow. As such, the correlation of consecutive shadowgrams enables seedless velocity measurements of the flow \citep{settles2006,hargather2011,settles2017}. In order for the pseudo-tracers to be distinguishable in consecutive images, thermal gradients should be sparsely distributed along the optical axis, and the diffusive timescale of the flow should far exceed its advective timescale. The applicability of shadow image velocimetry to turbulent flows of carbon dioxide at supercritical pressure in particular was previously demonstrated by \citet{okamoto2003}. Whereas Particle Image Velocimetry (PIV) can yield near-planar velocities, the integrating property of shadowgraphy yields measurements of eddy motion across the full width of the test section when using shadow image velocimetry. Much like planar PIV, shadow image velocimetry only yields displacements in directions perpendicular to the optical axis. In the current work, a fast-Fourier-transform-based cross-correlation algorithm is used to yield displacements between consecutive shadowgrams. An interrogation window size of 128$\times$128 pixels with 80$\%$ overlap is used to capture a sufficient amount of property-variant structures in each correlation window \citep{settles2006}.

\subsubsection{Background Oriented Schlieren (BOS)}
The magnitude of the vertical refractive index gradients across the test section can be evaluated by considering the deformation of a known pattern by the highly property-variant $\text{CO}_2$. In the current Background Oriented Schlieren (BOS) configuration, an in-focus background pattern is placed on the left side of the test section in cross section~\textbf{B} of figure~\ref{fig:schematic}. Here, a computer-generated and known dot pattern printed on a transparent plastic sheet is placed along the optical axis. By placing diffusing glass before the background pattern, the incoming light is decollimated. At the expense of the global image brightness, the diffuse light is found to produce a more homogeneously illuminated and globally sharp image when refractive index gradients are present in the test section channel. When a diffuser is used, the sensor exposure time is increased to 150 $\mu$s to provide sufficient illumination. Using the image correlation method previously used for shadow image velocimetry in this work, the apparent deformation $(\Delta x(x,y),\Delta y(x,y))$ of the background pattern in the image plane caused by gradients in refractive index in the test section can be measured. By assuming small deflection angles, the unidirectional, vertical deformation of the background pattern, and the presence of a schlieren object that is homogeneous across the full width of the channel, the path-integrated spatial refractive index gradient $\partial n/\partial y$ can be expressed as
\begin{equation}
    \frac{\partial n}{\partial y}=\frac{n_0 \Delta y}{MZ_{\text{D}}w},
\end{equation} \citep{schroder2009, raffel2015}. Here, $n_0$ is the reference refractive index of the carbon dioxide when it is homogeneous in density. Furthermore, $M$ is the magnification of the optical system, and $Z_{\text{D}}$ is the distance from the middle of the channel to the background pattern. Unfortunately, large density gradients are present throughout the limited field of view of the current setup, and the exact magnitude of the path-integrated density is unknown within the imaged volume. As such, no appropriate set of boundary conditions can be posed for the calculation of a line-of-sight integrated density field for the current configuration \citep{venkatakrishnan2004}. Nevertheless, the current BOS approach can be used for a relative comparison of density gradients in the channel, and for the evaluation of the applicability of particle-based optical diagnostics with the studied fluid.

\subsection{Experimental procedure}\label{sec:meth:procedure}
As carbon dioxide dilates strongly for the conditions and flow rates of the current study, all data is gathered by following a measurement procedure that aims to minimize thermal gradients within the test section block itself. The test section is initially at the ambient temperature. In order to increase its temperature to the desired value, a large mass flow rate of carbon dioxide at the appropriate temperature is forced through the thermally insulated test section for several hours. Over time, the measured temperature drop over the test section $(T_{\text{b1}}-T_{\text{b2}})$ reduces, as the temperature of the test section converges towards a steady state. When changing the inlet condition to reach different operating conditions, the procedure to reach steady state is repeated.

When the test section temperature reaches a value sufficiently close to the desired test section inlet temperature, therewith at a low value of $(T_{\text{b1}}-T_{\text{b2}})$, a measurement cycle is commenced. At a constant mass flow rate, the imposed foil heating rate is gradually increased in a stepwise manner. In order to maintain a constant pressure and mass flow rate $\dot{m}$, the applied heating rate to drive the natural circulation loop is gradually decreased. The acquisition of images and loop sensor data is started some time $t_{\text{w}}$ after imposing each step in heating power. In the current work, $t_{\text{w}}=120$ s. After the completion of all heating steps, the mass flow rate is changed, and the above process is repeated.
\section{Cases and parameter space}\label{sec:cases}
\begin{figure}
    \centering
\includegraphics[width=.65\textwidth]{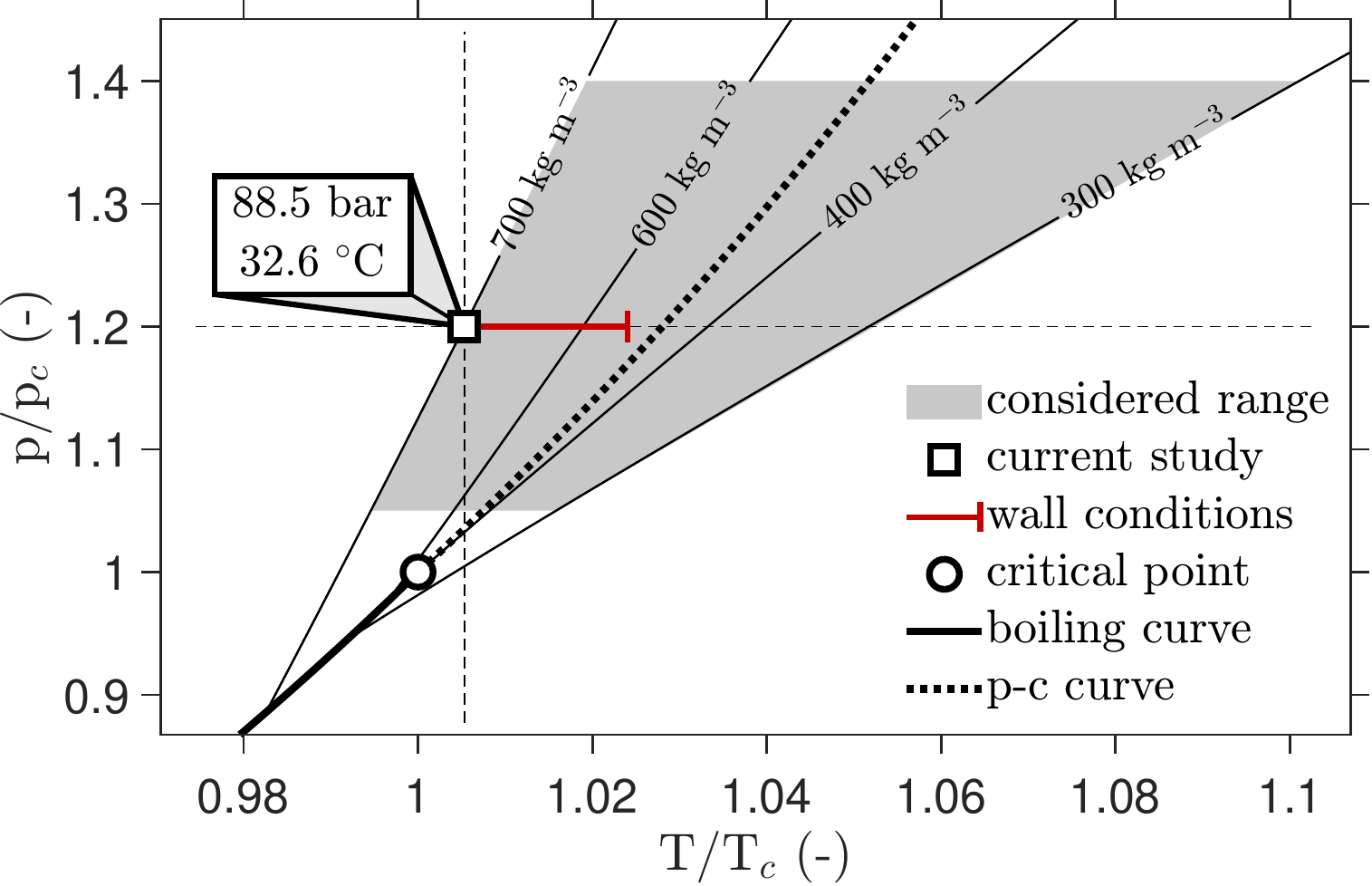}
\caption{Range of thermodynamic conditions of carbon dioxide at the test section inlet for the current study. Any presented results are evaluated at a nominal density and a nominal pressure of ${\rho=700\text{ kg m}^{-3}}$ and $p=1.2$ $p_{\text{c}}$, respectively. Here, $p_{\text{c}}$ is the critical pressure.}
\label{fig: ptplot}
\end{figure}
\subsection{Thermodynamic conditions}\label{sec:res:thermodynamic conditions}
All presented results are recorded at the same thermodynamic conditions at the test section inlet, whereas the imposed heating rates and the applied mass flow rates are varied throughout this work. The nominal thermodynamic state of the test section inlet ($T_{\text{b}1}$ and $P_{\text{b}1}$ in figure~\ref{fig:schematic}) of the current study is indicated in figure~\ref{fig: ptplot}. The full thermodynamic range that was initially explored is indicated in grey in figure~\ref{fig: ptplot}. 

Within the considered thermodynamic range, channel inlet conditions that are liquid-like provide the most distinguishable shadowgrams. Therefore, such conditions are chosen as the initial state for the current experiments. On the contrary, images of near-pseudo-critical and gas-like neutrally buoyant carbon dioxide reveal the intermittent presence of blurry, irregular, downward moving structures. Whereas downward plumes are similarly perceived when the $\text{CO}_2$ is denser, they are less prevalent, their presence is more sporadic and their outlines remain sharp in a shadowgram. These secondary flows are likely the result of the moderate cooling of the carbon dioxide by the ambient air, either through the insulated walls of the test section, or via the less insulated optical visors. Within the considered domain, the temperature of the carbon dioxide at the test section inlet exceeds the ambient temperature. On the right side of the pseudo-critical curve in figure~\ref{fig: ptplot}, the driving temperature difference with the surroundings is relatively large. Furthermore, the cooling of the gas-like medium brings about sharp changes in refractive index in the steep, non-linear near-pseudo-critical region. To the left of the p-c curve, both the driving temperature of the cooling by the ambient and the thermodynamic gradients induced by the cooling are significantly lower. Therewith, the moderately colder structures in the flow do not sufficiently distort the light through the channel to blur the resulting image in the liquid-like region of the considered thermodynamic parameter range. 

Additionally, the large near-pseudo-critical gradients in density that are induced when the liquid-like carbon dioxide is heated, make that strong stratifications can be reached with the attainable experimental heating rates. The range of measured surface temperatures $T_{\text{s}1}$ and $T_{\text{s}2}$ is indicated using the red horizontally bounded line in figure~\ref{fig: ptplot}. For the presented results, the pseudo-critical curve is not crossed in the near-wall region.
\begin{figure}
    \centering
\includegraphics[width=.95\textwidth]{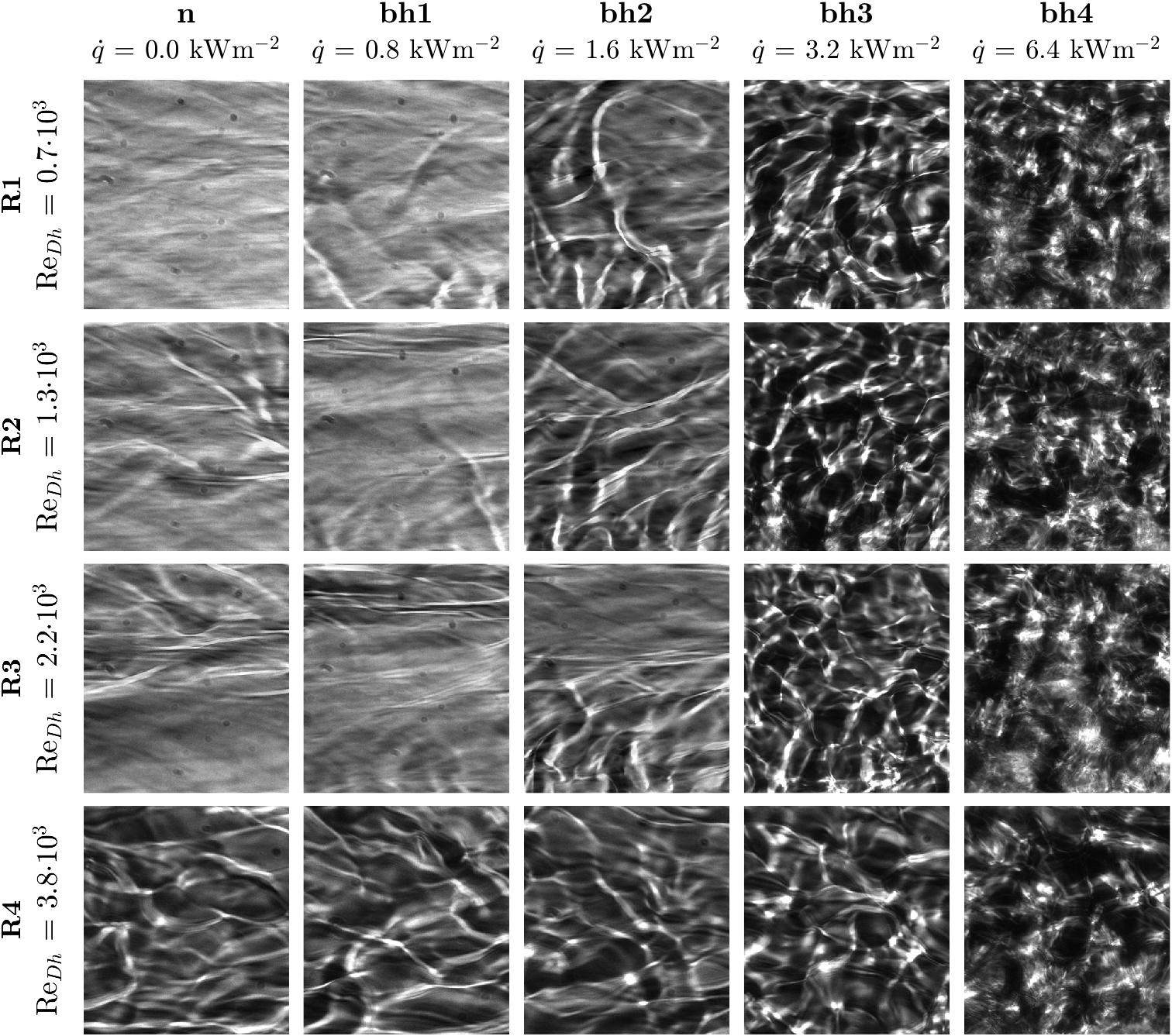}
 \caption{Instantaneous shadowgrams of bottom-heated channel flows of carbon dioxide at the thermodynamic conditions indicated in figure~\ref{fig: ptplot}. The relevant nominal Reynolds numbers $\Rey_{Dh}$ and the nominal imposed heating rates $\dot{q}$ are displayed on the vertical and the horizontal axes, respectively. The cases, e.g. R2:bh1, are labeled by an indicator of the nominal Reynolds number (\textbf{R}) and an indicator of the applied heating ($\textbf{n}$ or $\textbf{bh}$). Here, $\textbf{n}$ represents an unheated flow, whereas $\textbf{bh}$ implies that bottom-upward heating is applied. In all cases, the flow is from left to right.}
\label{fig:casebot}
\end{figure}
\begin{figure}
    \centering
\includegraphics[width=.95\textwidth]{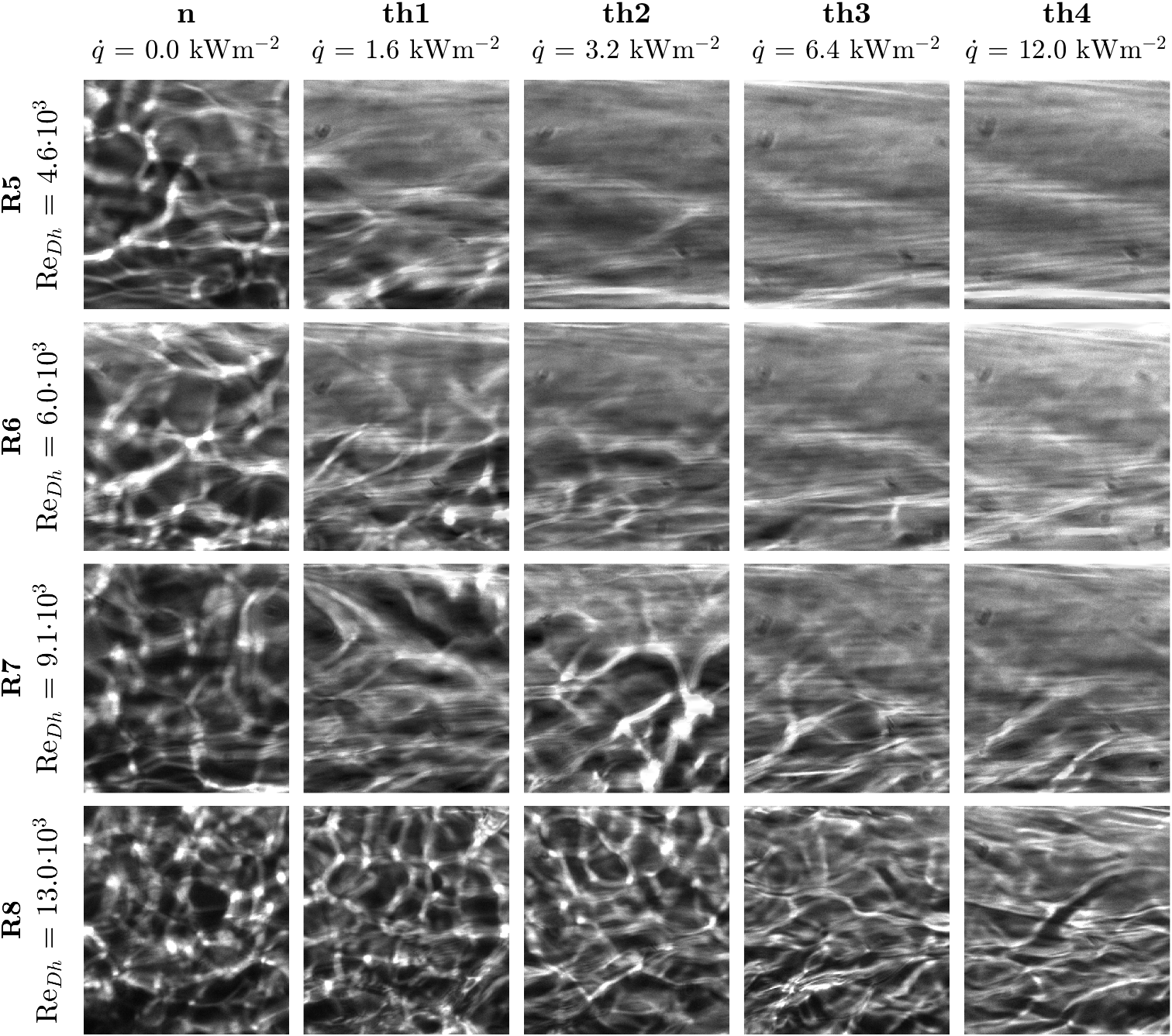}
\caption{Instantaneous shadowgrams of top-heated channel flows of carbon dioxide at the thermodynamic conditions indicated in figure~\ref{fig: ptplot}. The relevant nominal Reynolds numbers $\Rey_{Dh}$ and the nominal imposed heating rates $\dot{q}$ are displayed on the vertical and the horizontal axes, respectively The cases, e.g. R6:th1, are labeled by an indicator of the nominal Reynolds number (\textbf{R}) and an indicator of the applied heating ($\textbf{n}$ or $\textbf{th}$). Here, $\textbf{n}$ represents an unheated flow, whereas $\textbf{th}$ implies that top-downward heating is applied. In all cases, the flow is from left to right.}
\label{fig:casetop}
\end{figure} 
\subsection{Case overview}\label{sec:res:overview}
An overview of the cases considered in this work is given in figures \ref{fig:casebot} and \ref{fig:casetop}. Both figures show instantaneous shadowgrams of carbon dioxide at supercritical pressure. In figure~\ref{fig:casetop}, the carbon dioxide is heated in the top-down configuration shown in orange outlined cross section~\textbf{A} of figure~\ref{fig:schematic}. For the cases shown in figure~\ref{fig:casebot}, the test section is inverted and the bottom surface of the test section channel is heated. In both figures, the inlet mass flow rate $\dot{m}$ and the foil heating rate $\dot{q}$ are varied to yield twenty cases with varying hydraulic Reynolds numbers ${Re_{Dh}}$ and bulk Richardson numbers $Ri$ respectively, where
\begin{equation}  
    Re_{Dh}=\frac{\rho_{\text{b1}} U_{\text{imp}}D_{h}}{\mu_{\text{b1}}}, \: \: \:
    Ri=\frac{\rho_{\text{b1}}-\rho_{\text{w}}}{\rho_{\text{b1}}}\cdot\frac{gh}{U_{\text{imp}}^2}.
\end{equation}
\begin{table}
  \begin{center}
\def~{\hphantom{0}}
    \begin{tabular}{rrrrr}
        %\underline{figure~\ref{fig:casebot}}
         & \textbf{bh1} & \textbf{bh2} & \textbf{bh3} & \textbf{bh4} \vspace{0.15cm} \\
 \textbf{R1}&$12.64 \pm2.42$&$15.53 \pm3.18$&$16.13 \pm4.03$&$19.48 \pm5.57$ \\
 \textbf{R2}&$3.33 \pm1.80$&$6.41 \pm1.85$&$12.13 \pm1.96$&$14.91 \pm1.98$ \\
 \textbf{R3}&$4.13 \pm0.89$&$4.96 \pm0.69$&$7.90 \pm0.91$&$9.69 \pm0.67$ \\
 \textbf{R4}&$0.86 \pm0.45$&$1.66 \pm0.68$&$2.62 \pm0.37$&$3.58 \pm0.37$ \\
        \multicolumn{5}{c}{} \\
        %\underline{figure~\ref{fig:casetop}}
         & \textbf{th1} & \textbf{th2} & \textbf{th3} & \textbf{th4} \vspace{0.15cm} \\
 \textbf{R5}&$1.49 \pm0.23$&$2.38 \pm0.15$&$5.28 \pm0.23$&$11.51 \pm0.34$ \\
 \textbf{R6}&$0.78 \pm0.14$&$1.19 \pm0.09$&$2.75 \pm0.17$&$6.69 \pm0.24$ \\
 \textbf{R7}&$0.35 \pm0.07$&$0.61 \pm0.06$&$1.05 \pm0.07$&$2.54 \pm0.11$ \\
 \textbf{R8}&$0.15 \pm0.03$&$0.26 \pm0.03$&$0.48 \pm0.02$&$1.26 \pm0.05$ \\
    \end{tabular} 
  \caption{Values of $Ri$ for the non-neutrally buoyant snapshots shown in figures \ref{fig:casebot} and \ref{fig:casetop}. The cases, e.g. R1:bh1, are labeled by an indicator of the nominal Reynolds number (\textbf{R}) and an indicator of the applied heating ($\textbf{bh}$ or $\textbf{th}$). Here, $\textbf{bh}$ represents a flow to which bottom-upward heating is applied, whereas $\textbf{th}$ indicates that top-downward heating is applied. The data is shown with intervals of $\pm 2\sigma$.}
  \label{tab:Ricases}
  \end{center}
\end{table}
\begin{figure}
    \centering
    \begin{subfigure}[t]{0.5236\textwidth}
         \centering
         \includegraphics[width=\textwidth]{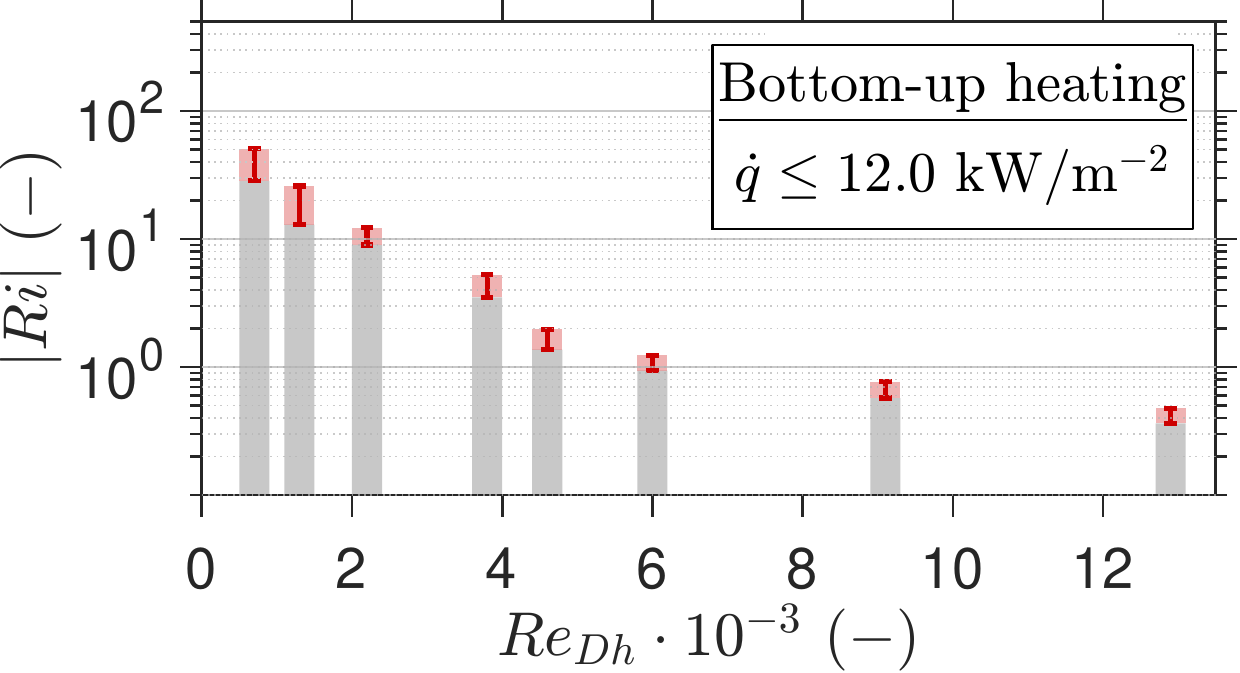}
         %\label{fig: SSnoheat0}
    \end{subfigure}
    \hfill
        \begin{subfigure}[t]{0.4618\textwidth}
         \centering
         \includegraphics[width=\textwidth]{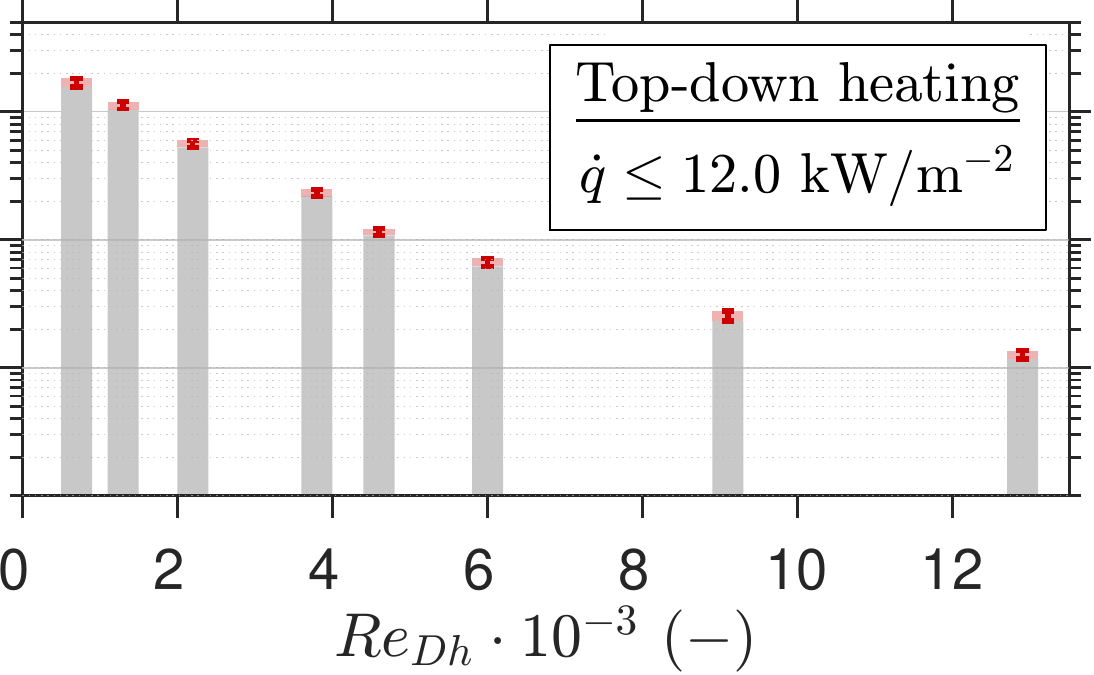}
         %\label{fig: SSnoheat1}
    \end{subfigure}
    \caption{Range of $Ri$ considered in this work for the nominal Reynolds numbers $Re_{\text{Dh}}$ indicated in figures \ref{fig:casebot} and \ref{fig:casetop}. The carbon dioxide is heated in the unstably stratified configuration on the left, whereas a stably stratified configuration is considered for the data on the right. The errorbars correspond to $\pm 2\sigma_{Ri}$ of the time series with the largest value of $Ri$.}
    \label{fig: Ri-Re}
\end{figure}
Here, the inlet density $\rho_{\text{b1}}$ and the inlet viscosity $\mu_{\text{b1}}$ are evaluated using $T_{\text{b1}}$ and $P_{\text{b1}}$ of figure~\ref{fig:schematic}. Thereafter, $\rho_{\text{b1}}$ is used to evaluate the imposed bulk velocity $U_{\text{imp}}$, with
\begin{equation}
    U_{\text{imp}}=
    \frac{\dot{m}}{\rho_{\text{b1}}A_{\text{cs}}}. 
\end{equation}
Here, the test section cross-sectional flow area is defined as $A_{\text{cs}}=hw$. The bulk Richardson number is evaluated using wall density $\rho_{\text{w}}$. Under the assumption of constant test section pressure, $\rho_{\text{w}}$ is evaluated at $P_{\text{b1}}$ and the mean wall temperature $T_{\text{w}}$, where
\begin{equation}
T_{\text{w}}=\frac{1}{2}\left(T_{\text{s1}}+T_{\text{s2}}\right).
\end{equation}
The corresponding values of $Ri$ for the non-neutrally buoyant cases shown in figures \ref{fig:casebot} and \ref{fig:casetop} are given in table \ref{tab:Ricases}. The complete range of $Ri$ obtained with the available surface heating power is presented in figure~\ref{fig: Ri-Re}. As the hydraulic Reynolds number $Re_{Dh}$ is increased, the attained value of $Ri$ decreases. Furthermore, at a constant heating rate, the measured value of $Ri$ is consistently higher for the top-down heated configuration than it is for bottom-up heating.
\section{Results}\label{sec:results}
\subsection{Characterisation of unheated base flow}
\begin{figure}
    \centering
    \begin{subfigure}[t]{0.2038\textwidth}
         \centering
         \includegraphics[width=\textwidth]{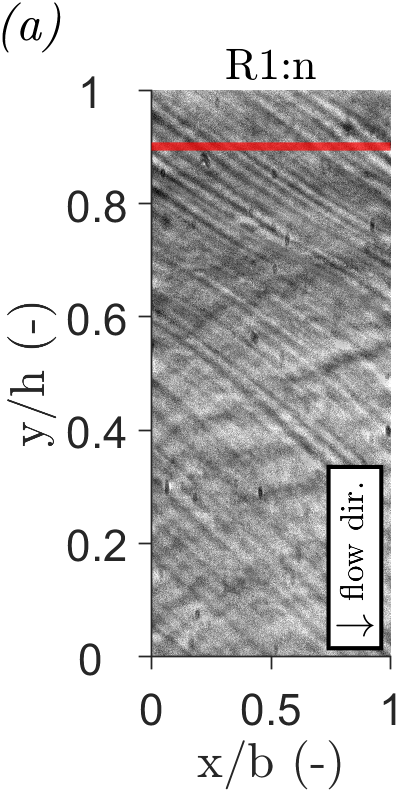}
         %\label{fig: SSnoheat0}
    \end{subfigure}
    \hfill
        \begin{subfigure}[t]{0.2033\textwidth}
         \centering
         \includegraphics[width=\textwidth]{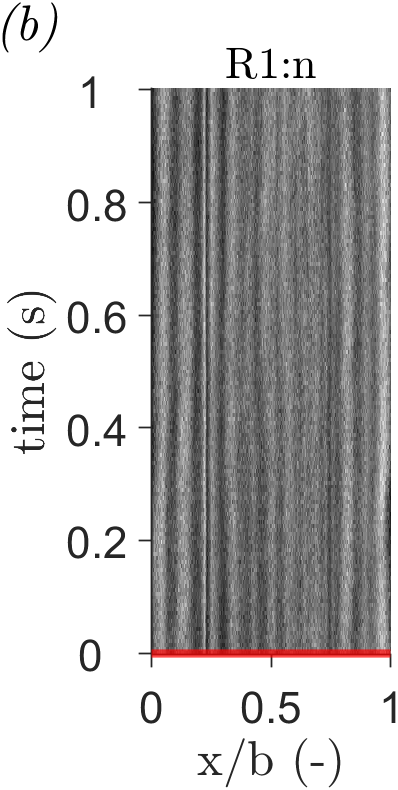}
         %\label{fig: SSnoheat1}
    \end{subfigure}
    \hfill
            \begin{subfigure}[t]{0.1335\textwidth}
         \centering
         \includegraphics[width=\textwidth]{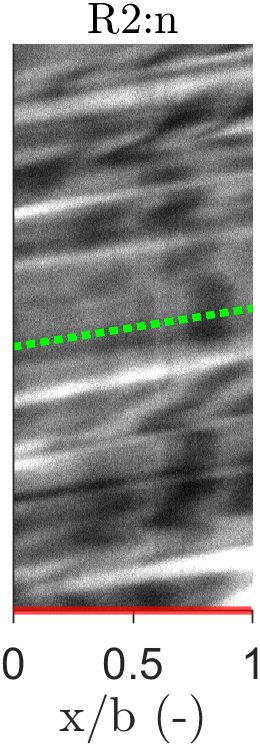}
         %\label{fig: SSnoheat2}
    \end{subfigure}
    \hfill
            \begin{subfigure}[t]{0.1335\textwidth}
         \centering
         \includegraphics[width=\textwidth]{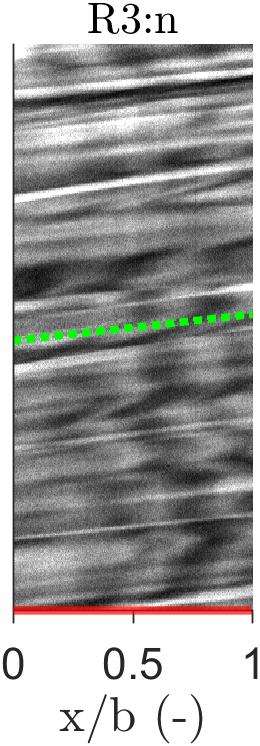}
         %\label{fig: SSnoheat3}
    \end{subfigure}
    \hfill
            \begin{subfigure}[t]{0.1335\textwidth}
         \centering
         \includegraphics[width=\textwidth]{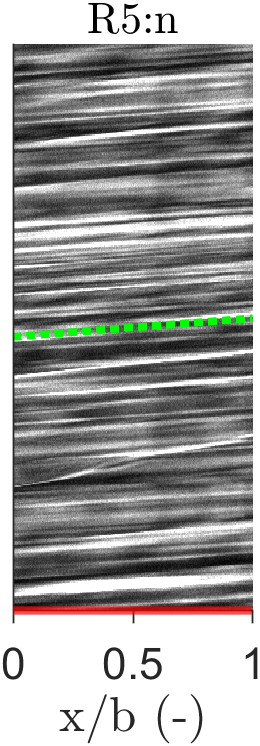}
         %\label{fig: SSnoheat4}
    \end{subfigure}
    \hfill
            \begin{subfigure}[t]{0.1335\textwidth}
         \centering
         \includegraphics[width=\textwidth]{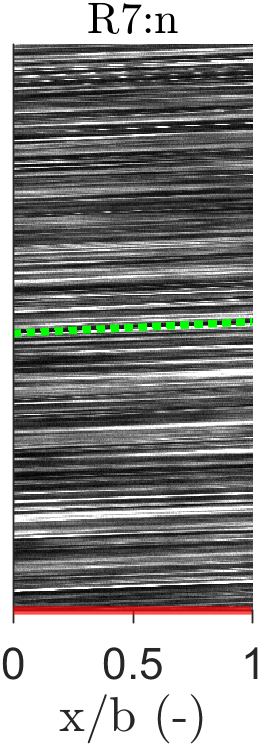}
         %\label{fig: SSnoheat5}
    \end{subfigure}
    \hfill
    \caption{Space-time (b) representation of the optical signal at the horizontal line that is indicated in the instantaneous shadowgram shown in (a). The value of $\Rey_{Dh}$ is progressively increased by increasing the mass flow rate for the cases shown in (b).}
    \label{fig: SSnoheat}
\end{figure}

For the conditions considered in this work, structures that are naturally present in the channel can be perceived in the shadowgrams and persist throughout consecutive frames. The shadowgraphy registers variations of the (second derivative of the) refractive index in the carbon dioxide throughout the channel, resulting from the turbulent mixing of the thermally heterogeneous fluid. In the shadowgrams, the variations manifest as fluctuations in the image brightness. The fluctuations maintain a consistent shape throughout consecutive frames, indicating that their timescale of mixing exceeds the interval between consecutive recordings. Therefore, their movement can be distinguished. The axial evolution of the fluctuations is shown in figure~\ref{fig: SSnoheat}. Here, the shadowgrams shown in the lefmost columns of figures \ref{fig:casebot} and \ref{fig:casetop} are considered, in which the $\text{CO}_2$ is neutrally buoyant. In figure~\ref{fig: SSnoheat}, the parameter $b$ describes the achieved image width for the current frame rates. The optical signal of the shadowgraphy is shown over time at the red horizontal line of the snapshot that is indicated in figure~\ref{fig: SSnoheat}a. The resulting space-time graphs show the passage of thermal structures in the form of angled streaks of constant brightness. The nominal angle of the streaks in the space-time graphs is indicated by green dashed lines for each case. Given that no structures are perceived over time along the red curve for case R1:n, a nominal angle cannot be inferred from the graph. As the imposed Reynolds number is increased for the other cases, the observed structures pass the red line in a shorter time span, and the angle of the streaks with the horizontal axis decreases. Evidently, a velocity-like quantity can be deducted from the streak angles of passing thermal structures in the current representation.
\begin{figure}
    \centering
    \hspace{1.4cm}
    \begin{subfigure}[t]{0.42\textwidth}
         \centering
         \includegraphics[width=\textwidth]{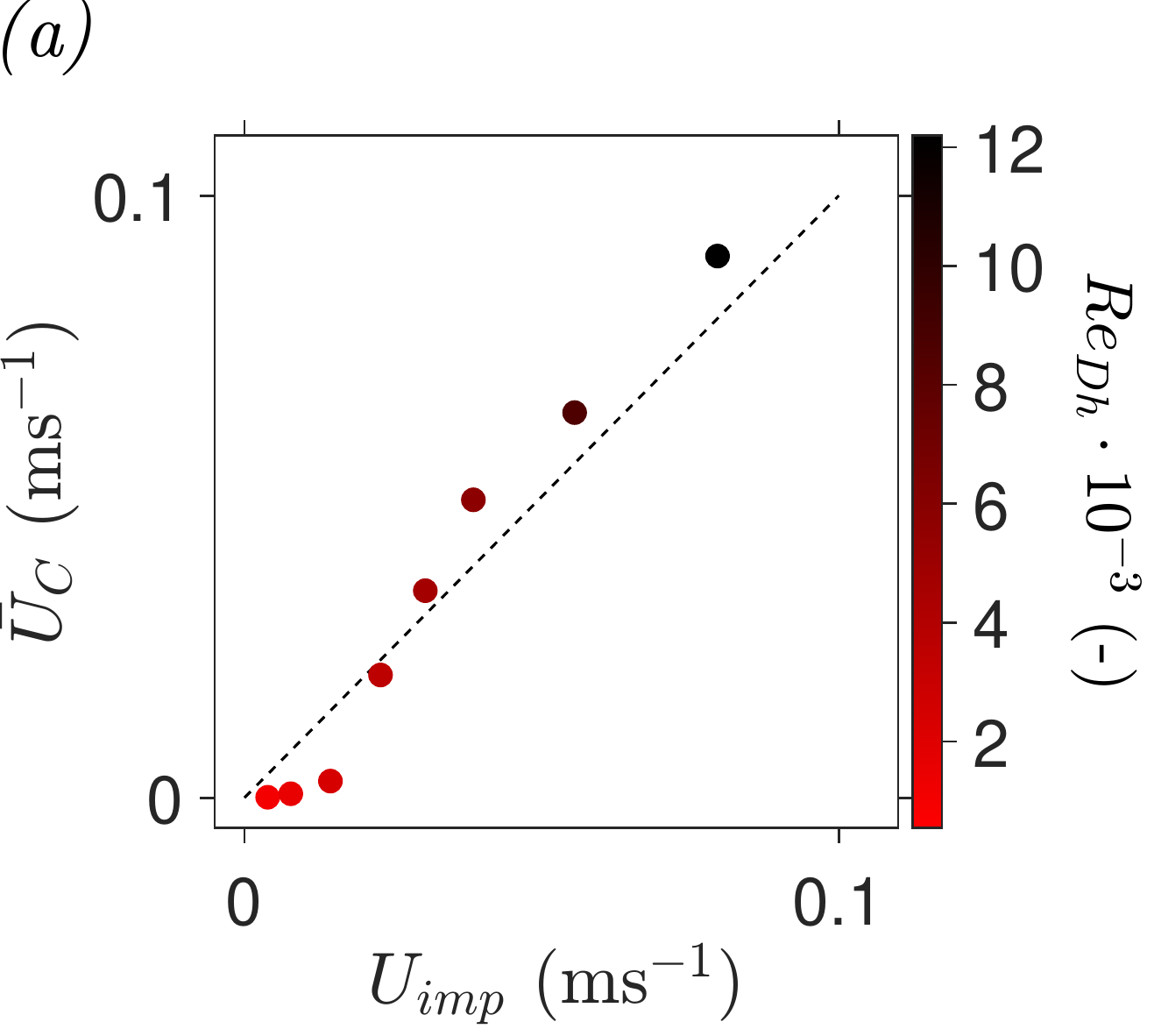}
         %\label{fig: piv1}
    \end{subfigure}
    \hspace{0.4 cm}
    \begin{subfigure}[t]{0.2918\textwidth}
         \centering
         \includegraphics[width=\textwidth]{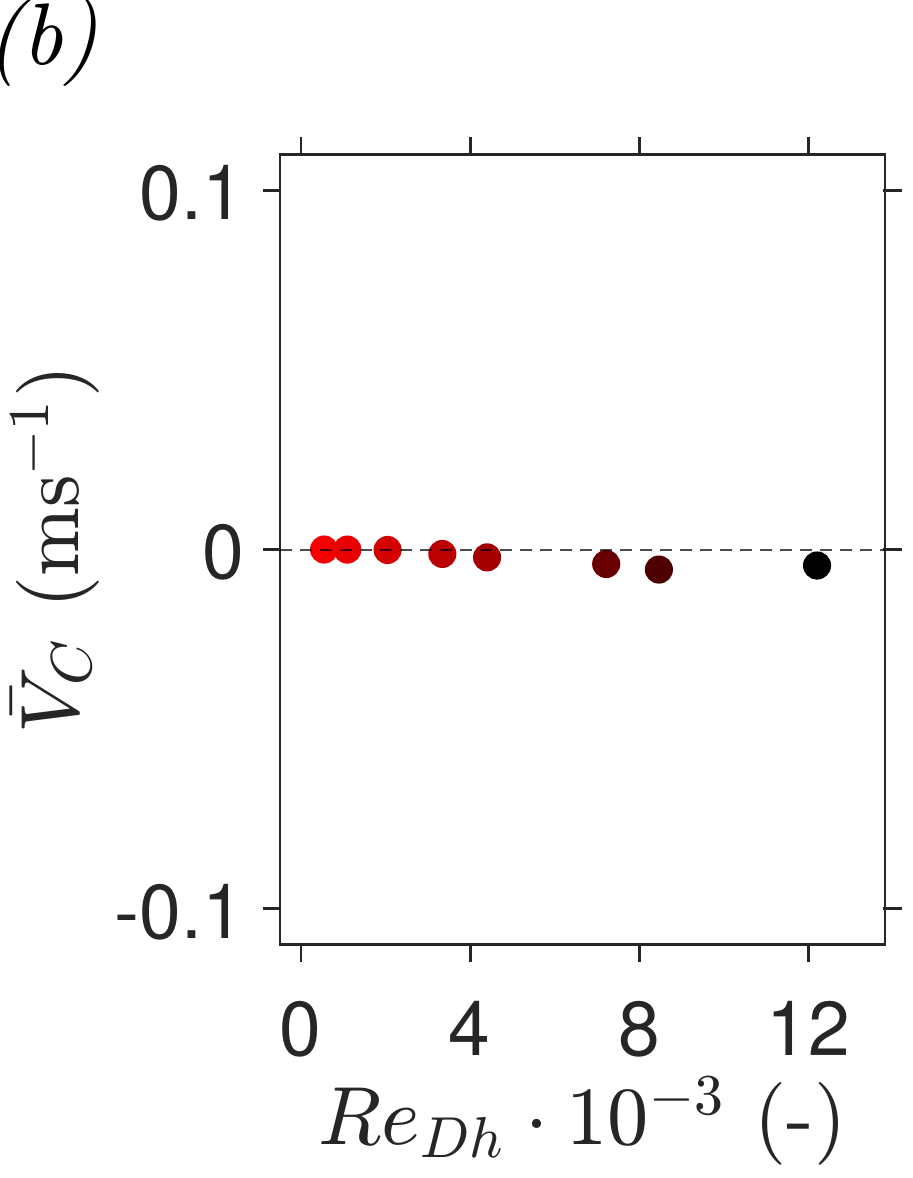}
         %\label{fig: piv2}
    \end{subfigure}
    \hfill
    \vspace{-0.1cm}
    \begin{subfigure}[t]{0.9\textwidth}
         \centering
         \includegraphics[width=\textwidth]{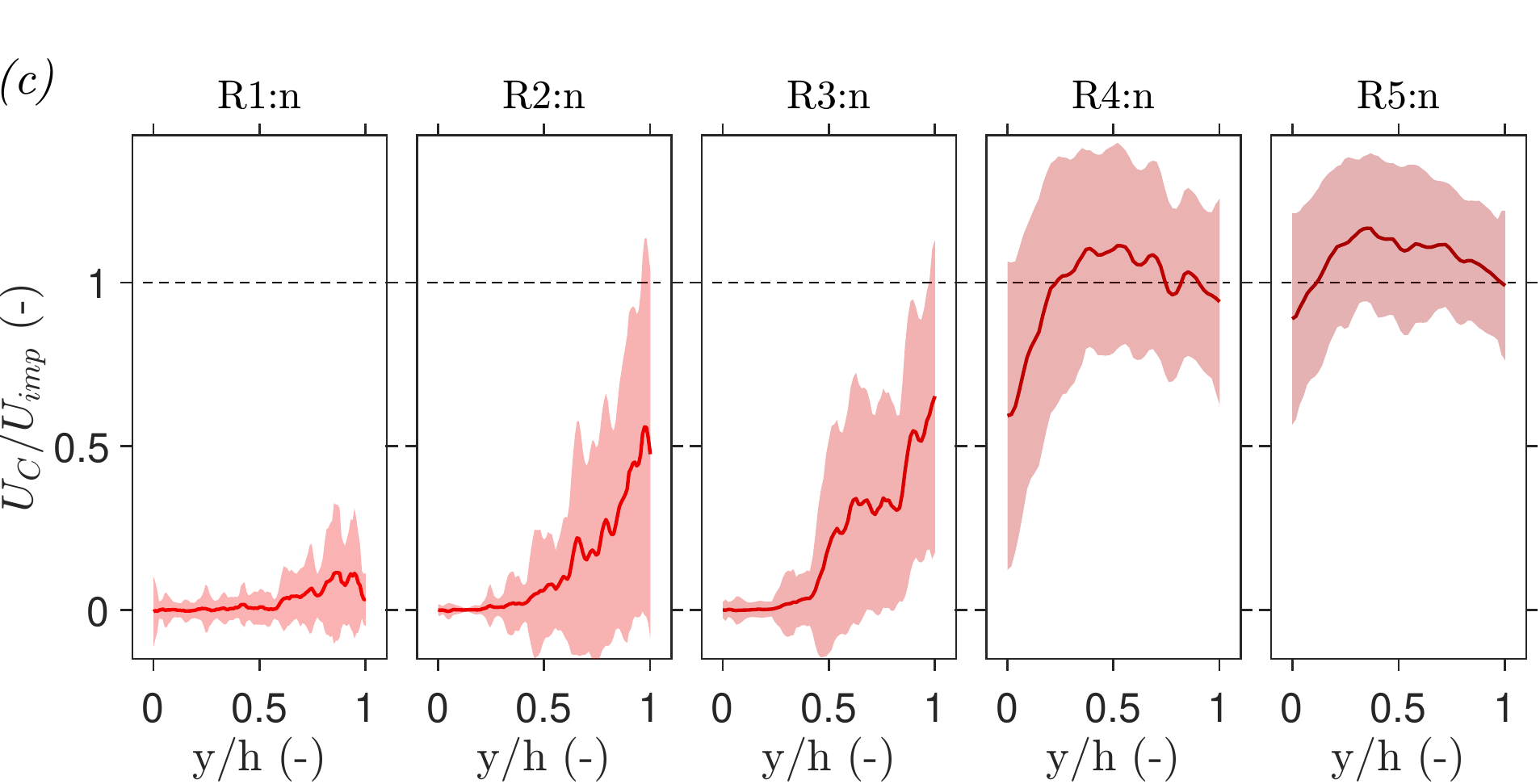}
         %\label{fig: piv3}
    \end{subfigure}
\caption{Comparison of the correlated velocity with the imposed velocity. Subfigure (a) shows the correlated mean velocity $\bar{U}_{C}$ with respect to the measured velocity. Subfigure (b) shows the mean value of the vertical component $\bar{V}_{C}$. A comparison of several instantaneous profiles of $U_{C}$ with the measured bulk velocity is shown in (c). The data in (a) and (b) are shown with errorbars of $\pm 2\sigma/\sqrt{N}$, where $N$ is the number of frame pairs. The data in (c) contains errorbars of $\pm\sigma$.}
\label{fig:pivcalibration}
\end{figure}

In order to determine whether the movement of the image brightness fluctuations corresponds to the movement of the fluid, the correlated velocity of the fluctuations $U_{C}$ is measured and subsequently compared to the the velocity imposed by the natural circulation loop $U_{\text{imp}}$. In order to measure $U_{C}$, consecutive shadowgrams are correlated to yield the instantaneous displacements of the distinctly shaped image brightness fluctuations in all directions perpendicular to the optical axis. Here, the total image height is assumed to correspond to $h$. Furthermore, the averaged image is assumed to correspond linearly to the object plane when the $\text{CO}_2$ remains unheated. As such, the means of image-plane displacements $\Delta x$ and $\Delta y$ can be divided by the optical magnification $M$ to yield the physical displacements within the test section. Therewith, $U_{C}$ can be approximated using
\begin{equation}
    U_{C}=\frac{\Delta x}{M\Delta t},
\end{equation}
where $\Delta t$ is equal to the period between consecutive frames. However, when the carbon dioxide is heated, the averaged image deforms as a result of large local gradients in refractive index. Therefore, the imaged displacements no longer linearly correspond to the physical displacements as $M$ is no longer constant throughout the image, and an accurate average value of $U_{C}$ can no longer be directly deducted. As such, only image-plane displacements are reported for the cases with heating.

The correlated velocity of the refractive structures corresponds conditionally to the imposed velocity of the system. Figure~\ref{fig:pivcalibration} shows a comparison of the correlated velocities of the structures in the shadowgrams with the velocities of the natural circulation for unheated flows. In Figure~\ref{fig:pivcalibration}a, the correlated mean $x$-direction velocity $U_{C}$ (deducted from 2999 frame pairs) is compared to $U_{\text{imp}}$. The magnitude of the mean reconstructed velocity $\bar{U}_{C}$ corresponds to the actual bulk velocity $U_{\text{imp}}$ for sufficiently large Reynolds numbers, i.e. $Re_{Dh}\geq 3\times 10^{3}$. As shown in the two rightmost plots of figure~\ref{fig:pivcalibration}c, a mean velocity profile of $U_{C}$ is reconstructed in which the center-line velocity exceeds the correlated velocity near the channel walls. Furthermore, as expected for a hydrodynamically developed flow, the center-line velocity exceeds the bulk velocity of the channel. As such, the imaged structures are assumed to travel at the local velocity of the flow. Hence, the fluctuations in image brightness can serve as `pseudo-tracers' of the flow field when it is turbulent for the current thermodynamic conditions. However, the integrating property of shadowgraphy makes that variations of refractive index at any lateral location in the channel are imaged. Therefore, the resulting instantaneous displacement fields do not directly correspond to instantaneous, planar flow fields. Instead, the movement of all refractive structures throughout the channel is superimposed. Therefore, as the streamwise velocity of the pseudo-tracers varies laterally, their image velocimetry is characterized by relatively large fluctuations.

\begin{figure}
    \centering
    \begin{subfigure}[t]{0.2089\textwidth}
         \centering
         \includegraphics[width=\textwidth]{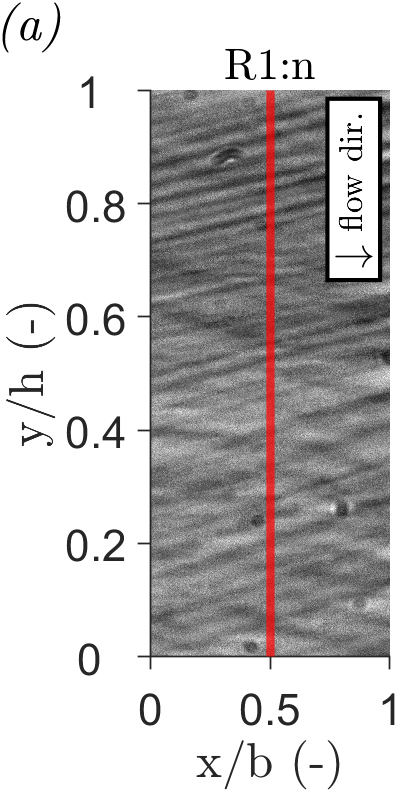}
         %\label{fig: SSnoheatv0}
    \end{subfigure}
    \hfill
        \begin{subfigure}[t]{0.2051\textwidth}
         \centering
         \includegraphics[width=\textwidth]{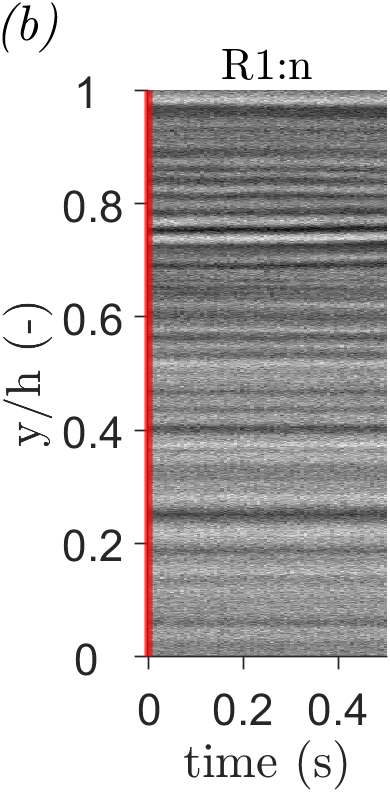}
         %\label{fig: SSnoheatv1}
    \end{subfigure}
    \hfill
            \begin{subfigure}[t]{0.134\textwidth}
         \centering
          \includegraphics[width=\textwidth]{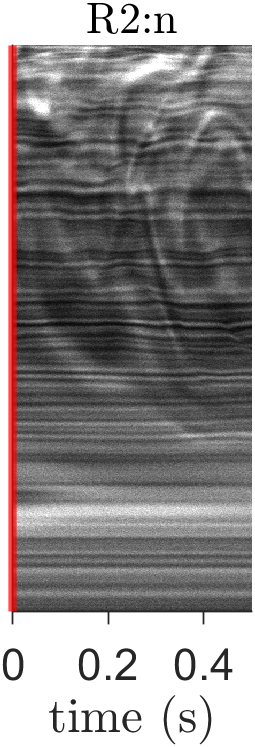}
         %\label{fig: SSnoheatv2}
    \end{subfigure}
    \hfill
            \begin{subfigure}[t]{0.134\textwidth}
         \centering
         \includegraphics[width=\textwidth]{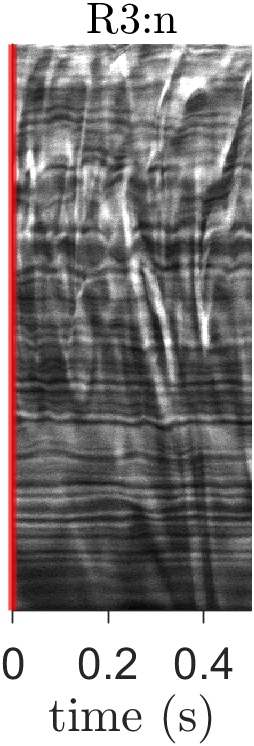}
         %\label{fig: SSnoheatv3}
    \end{subfigure}
    \hfill
            \begin{subfigure}[t]{0.134\textwidth}
         \centering
         \includegraphics[width=\textwidth]{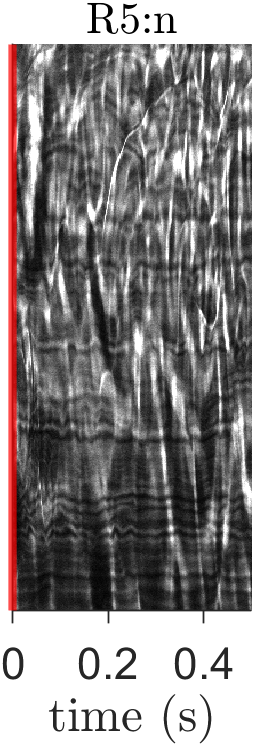}
         %\label{fig: SSnoheatv4}
    \end{subfigure}
    \hfill
            \begin{subfigure}[t]{0.134\textwidth}
         \centering
         \includegraphics[width=\textwidth]{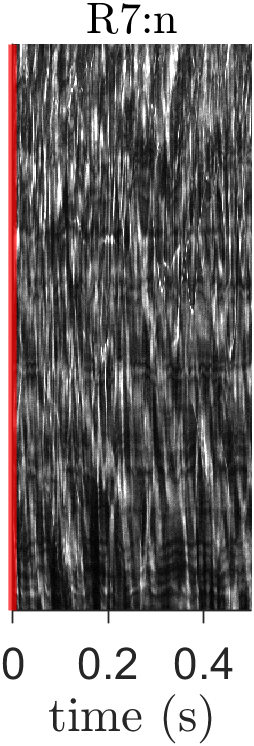}
         %\label{fig: SSnoheatv5}
    \end{subfigure}
    \hfill
    \caption{Space-time (b) representation of the optical signal at the vertical line that is indicated in the instantaneous shadowgrams shown in (a). The value of $\Rey_{Dh}$ is progressively increased by increasing the mass flow rate for the cases shown in (b).}
    \label{fig: SSnoheatv}
\end{figure}

At low velocities the structures may decorrelate, yielding no contribution to the displacement correlation, and therewith biasing the result to lower displacements. Figure~\ref{fig:pivcalibration}c shows this under-reconstruction of the bulk velocity in detail. In the figure, a range of reconstructed mean profiles of $U_{C}$ is presented, normalised with $U_{\text{imp}}$. NCL velocity $U_{\text{imp}}$ is increasingly under-captured for $Re_{Dh}\leq 3\times 10^3$, most notably so near the bottom channel surface. Figure~\ref{fig: SSnoheatv} elucidates the cause of this inadequate characterization of the channel bulk velocity. Here, figure~\ref{fig: SSnoheatv}b shows the optical data captured over time along the vertical red line in figure~\ref{fig: SSnoheatv}a. In the figure, strong fluctuations in image brightness are shown throughout the channel for large Reynolds number cases R5:n and R7:n, indicative of the passage of pseudo-tracers. As the Reynolds number is decreased, the flow becomes more sparsely 'seeded.' Initially, for R3:n and R2:n, fluctuations of the optical signal only persist in the region confined to the upper wall. In these cases the channel flow is likely thermally stratified, in which a separate layer that is lighter and more turbulent is formed on top of a denser, laminar layer. In figure~\ref{fig: SSnoheatv}b, the latter is characterized by an optical signal that is predominantly constant in brightness over time. Whereas this quiescent region is confined to the bottom wall for cases R3:n and R2:n, it spans the full channel height for case R1:n. As such, the channel flow in case R1:n is assumed to be entirely laminar. As the laminar layer is largely homogeneous in density, its shadowgraphy will not portray the passage of compressible structures. As a result, the correlation of consecutive images underestimates the actual motion of the fluid. Therefore, only where sufficient turbulent mixing is present in cases R2:n and R3:n, near the top boundary of the channel, a velocity in the order of magnitude of $U_{C}$ is captured in figure~\ref{fig:pivcalibration}c. As the flow is sufficiently mixed such that no stratified layers appear throughout the channel for cases R4:n and beyond, a more appropriate mean flow field is captured throughout the channel.

Lastly, the characterization of the mean flow field shows a moderately negative mean $y$-direction velocity $\bar{V}_{C}$ for the larger values of $Re_{Dh}$ within the considered range. Here, the downward motion of sporadic cold plumes within the channel is captured by the image velocimetry. The downdraughts, likely caused by the slight cooling of the $\text{CO}_2$ by the ambient in the vicinity of the visors, have previously been discussed.

The visualisation of the flow field using thermal turbulence as tracers has revealed a mean streamwise flow field characterised by features similar to those of hydrodynamically developed flows with sub-critical fluids. However, the large uncertainty of the current method, and its inability to reconstruct a flow field when the flow is in a thermally homogeneous laminar flow state makes its applicability limited. As such, the complete characterisation of instantaneous, planar flow fields at supercritical and thermally heterogeneous conditions requires additional and yet to be developed experimental techniques. Nevertheless, the shadowgrams and the velocimetry thereof can reveal instantaneous fluid motion and flow patterns (or the lack thereof) across the channel at high spatial and temporal frequencies when there is sufficient turbulent mixing in the property-variant working fluid. As such, shadowgrams reveal the transient, local phenomena underlying the greatly atypical heat transfer correlations reported in open experimental literature. Therefore, shadowgraphy is used to study the modulation of the channel flow by thermal stratification from here onwards.
\subsection{Unstable stratification}\label{sec:res:unstable}
\begin{figure}
    \centering
    \begin{subfigure}[t]{0.2093\textwidth}
         \centering
         \includegraphics[width=\textwidth]{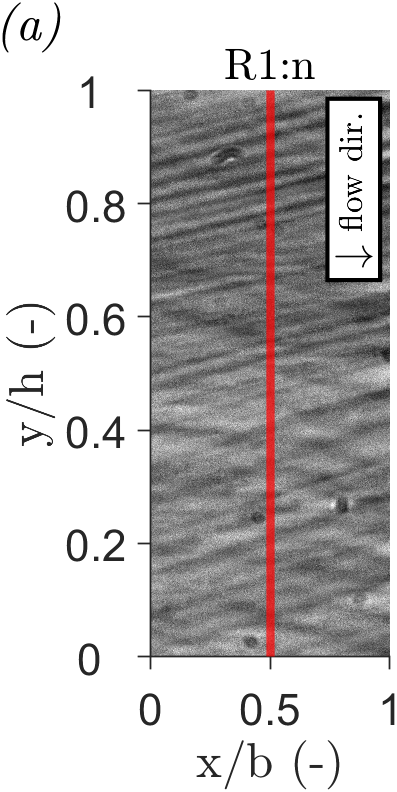}
         %\label{fig: SSbotheat0}
    \end{subfigure}
    \hfill
        \begin{subfigure}[t]{0.2055\textwidth}
         \centering
         \includegraphics[width=\textwidth]{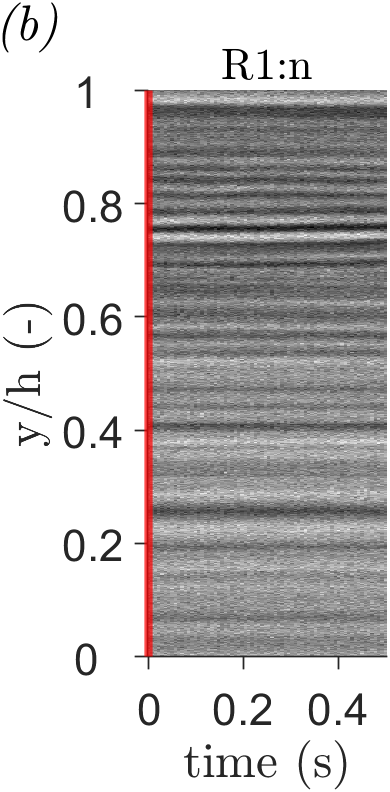}
         %\label{fig: SSbotheat1}
    \end{subfigure}
    \hfill
            \begin{subfigure}[t]{0.134\textwidth}
         \centering
         \includegraphics[width=\textwidth]{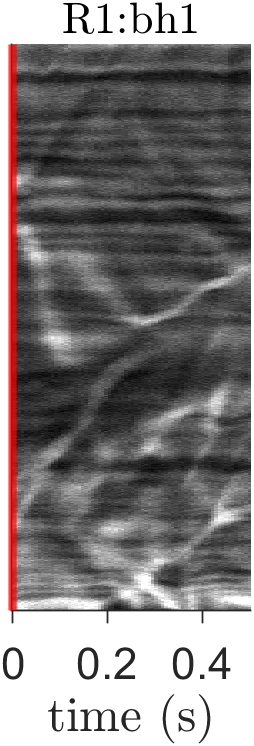}
         %\label{fig: SSbotheat2}
    \end{subfigure}
    \hfill
            \begin{subfigure}[t]{0.134\textwidth}
         \centering
         \includegraphics[width=\textwidth]{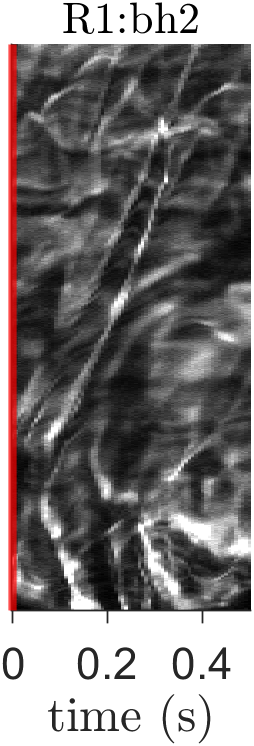}
         %\label{fig: SSbotheat3}
    \end{subfigure}
    \hfill
            \begin{subfigure}[t]{0.134\textwidth}
         \centering
         \includegraphics[width=\textwidth]{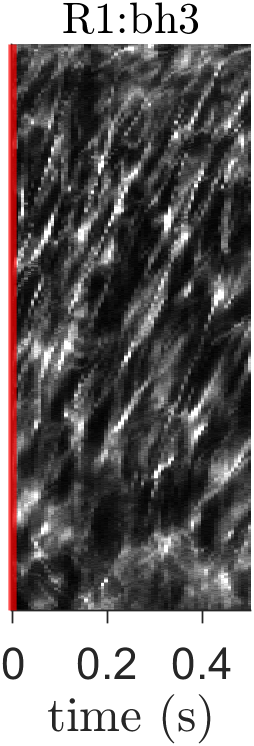}
         %\label{fig: SSbotheat4}
    \end{subfigure}
    \hfill
            \begin{subfigure}[t]{0.134\textwidth}
         \centering
         \includegraphics[width=\textwidth]{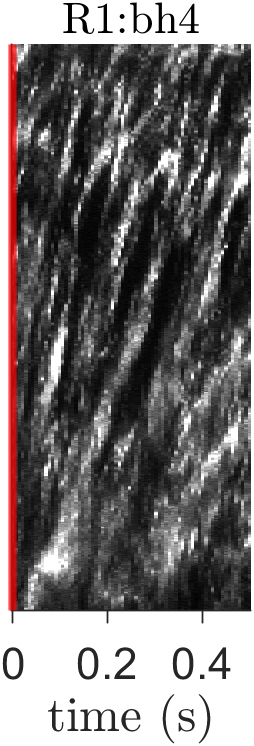}
         %\label{fig: SSbotheat5}
    \end{subfigure}
    \hfill
    \caption{Space-time (b) representation of the optical signal at the vertical line that is indicated in the instantaneous shadowgram shown in (a). The bottom-up heating rate is progressively increased for the consecutive cases shown in (b).}
    \label{fig: SSbotheat}
\end{figure}
\begin{figure}
    \centering
    \begin{subfigure}[t]{0.16\textwidth}
         \centering
         \includegraphics[width=\textwidth]{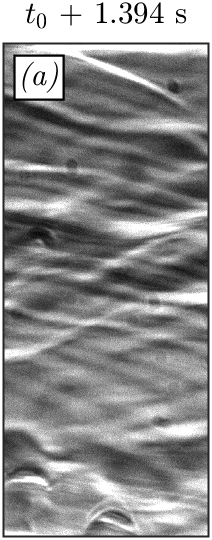}
         %\label{fig: filt1}
    \end{subfigure}
    \hfill
    \begin{subfigure}[t]{0.16\textwidth}
         \centering
         \includegraphics[width=\textwidth]{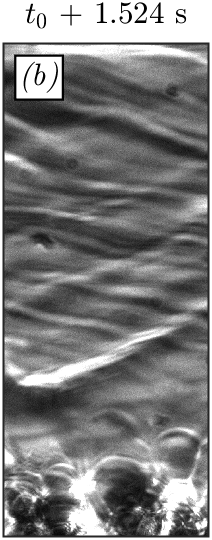}
         %\label{fig: filt2}
    \end{subfigure}
    \hfill
    \begin{subfigure}[t]{0.16\textwidth}
         \centering
         \includegraphics[width=\textwidth]{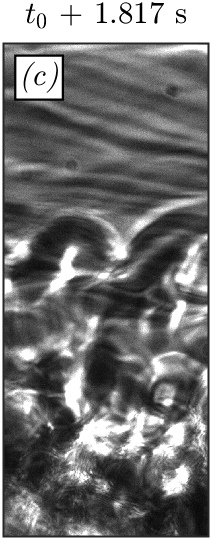}
         %\label{fig: filt3}
    \end{subfigure}
    \hfill
    \begin{subfigure}[t]{0.16\textwidth}
         \centering
         \includegraphics[width=\textwidth]{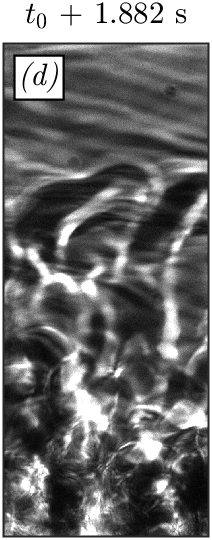}
         %\label{fig: filt4}
    \end{subfigure}
    \hfill
    \begin{subfigure}[t]{0.16\textwidth}
         \centering
         \includegraphics[width=\textwidth]{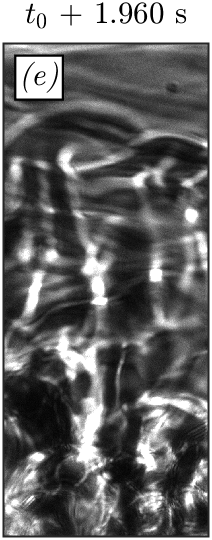}
         %\label{fig: filt5}
    \end{subfigure}
    \hfill
    \begin{subfigure}[t]{0.16\textwidth}
         \centering
         \includegraphics[width=\textwidth]{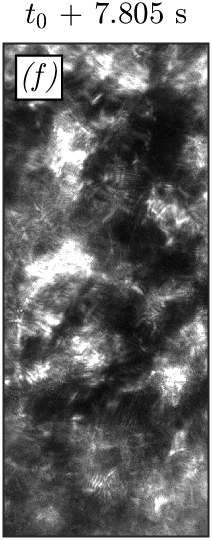}
         %\label{fig: filt6}
    \end{subfigure}
\caption{Series of shadowgrams that show the evolution of the flow after an instantaneous step in bottom-upwards heating from case R2:n to R2:bh4. At time $t_0$, the heating is commenced.}
\label{fig:plumes}
\end{figure}

When the rectangular channel is heated in the bottom-upward configuration, the perceived flow field within it is characterized by increased, mostly vertical movement. Currently, the cases shown in figure~\ref{fig:casebot} are considered. Figure~\ref{fig: SSbotheat} shows the evolution of the shadowgraphy signal of the vertical red line indicated in figure~\ref{fig: SSbotheat}a in figure~\ref{fig: SSbotheat}b. Here, the initially quiescent flow of case R1:n is disturbed by movement induced by the unstably stratified $\text{CO}_2$, and images of fluctuating brightness follow. Initially, for R1:bh1, the upward motion in the shadowgrams is predominantly confined to the near-wall region at the bottom of the channel. As warm, lighter plumes rise towards the top of the channel when buoyancy is moderate, they are advected by the bulk flow in the streamwise direction. For this process, the angle of a streak of constant brightness with respect to the vertical axis increases over time in a space-time graph. The process of plume advection by the mean flow is more clearly shown in figure~\ref{fig:plumes}. Here, the channel is imaged in the period after which the heating state is instantaneously changed from R2:n to R2:bh4. Mushroom-like plumes arise from the bottom of the channel in figures \ref{fig:plumes}a and \ref{fig:plumes}b, as part of a wide, turbulent front. The plumes at the top of the front are overturned in the streamwise direction of the flow as they near the top of the domain in figures \ref{fig:plumes}c-e. As such, they form rollers that are ordered one after another in the streamwise direction of the flow, similar to those simulated by \citet{garai2014} and \citet{pirozzoli2017} for mixed unstable convection in ideal fluids. For the conditions in case R1:bh1, the rollers are exclusively confined to the bottom two-thirds of the channel.

\begin{figure}
    \centering
    \begin{subfigure}[t]{0.241\textwidth}
         \centering
         \includegraphics[width=\textwidth]{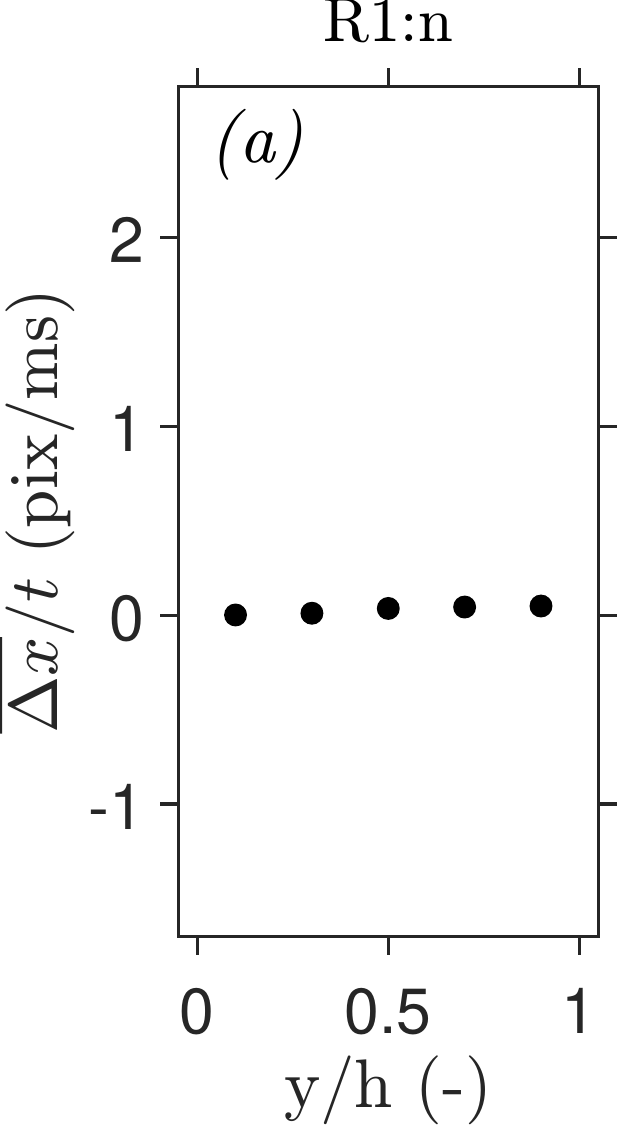}
         %\label{fig: dispbh1}
    \end{subfigure}
    \hfill
        \begin{subfigure}[t]{0.178\textwidth}
         \centering
         \includegraphics[width=\textwidth]{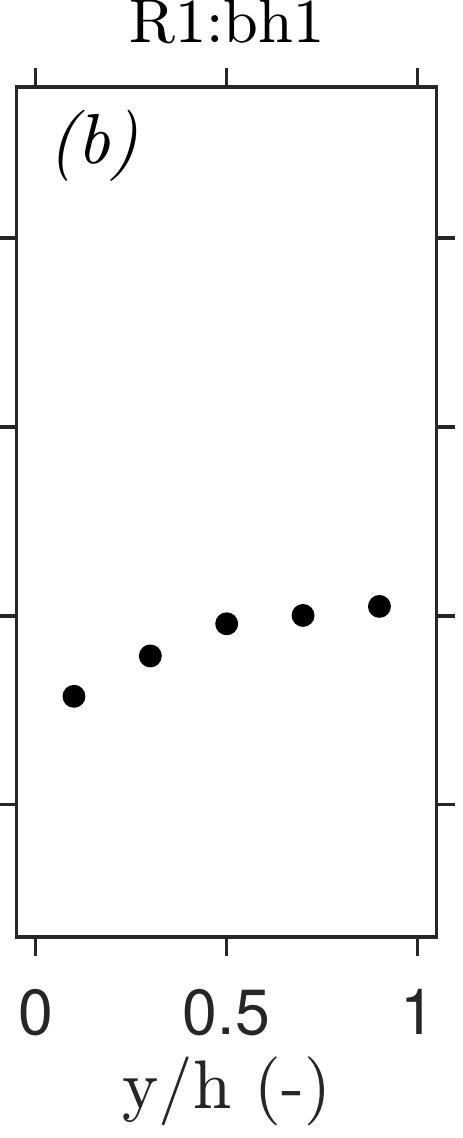}
         %\label{fig: dispbh2}
    \end{subfigure}
    \hfill
            \begin{subfigure}[t]{0.178\textwidth}
         \centering
         \includegraphics[width=\textwidth]{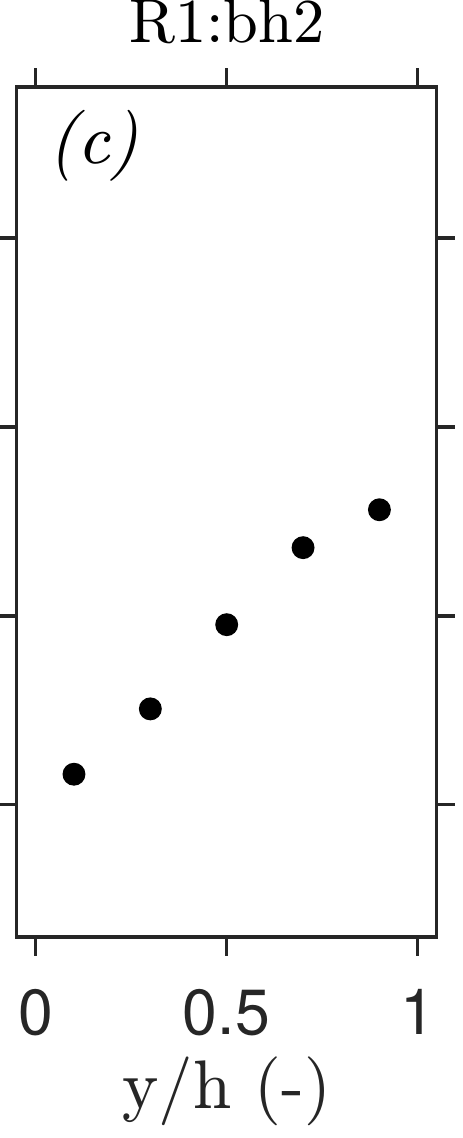}
         %\label{fig: dispbh3}
    \end{subfigure}
    \hfill
            \begin{subfigure}[t]{0.178\textwidth}
         \centering
         \includegraphics[width=\textwidth]{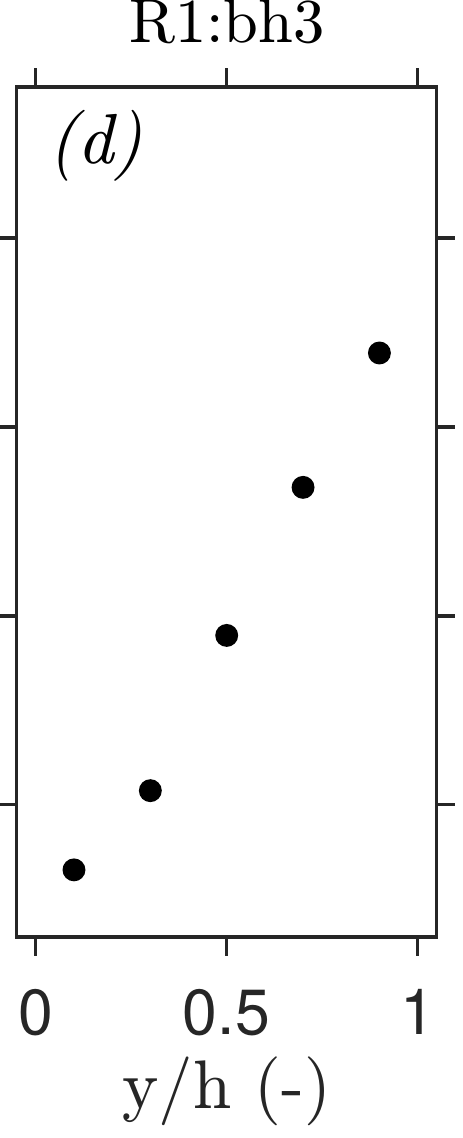}
         %\label{fig: dispbh4}
    \end{subfigure}
    \hfill
            \begin{subfigure}[t]{0.178\textwidth}
         \centering
         \includegraphics[width=\textwidth]{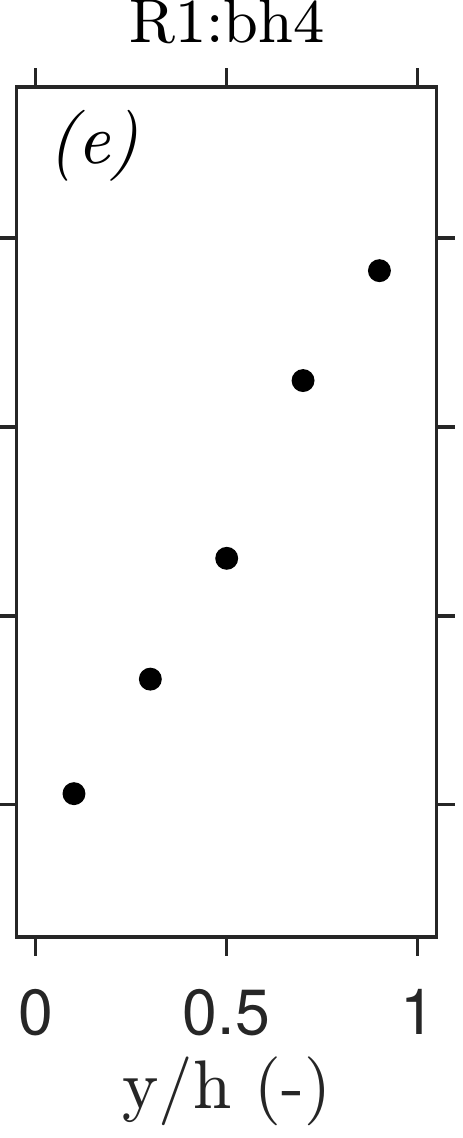}
         %\label{fig: dispbh5}
    \end{subfigure}
    \caption{Mean perceived horizontal displacement rates $\overline{\Delta x}/t$ for case $R1$, with progressively increased bottom-up heating rates in the consecutive subfigures. 
    % The data is shown with errorbars of $\pm 2\sigma/\sqrt{N}$, where $N$ is the number of frame pairs.
    }
    \label{fig: dispbh_U}
\end{figure}

As the heating rate is increased from R1:bh1 to R1:bh2, the thermal structures originating from the heated bottom wall sporadically reach the top surface of the channel. The movement of these updrafts is characterized by straight lines of constant brightness at an angle with the vertical axis in figure~\ref{fig: SSbotheat}b. As the heating rate is further increased for cases R1:bh3 and R1:bh4, the relative amount of plumes that traverse the channel in its entirety increases, and the image is eventually dominated entirely by the presence of these upward moving structures. Whereas shadowgraphy yields images of mostly constant brightness (see figure~\ref{fig: SSbotheat}a) when no heating is applied, instantaneous shadowgrams show a chaotic superposition of many plumes for cases R1:bh3 and R1:bh4, as shown in figure~\ref{fig:plumes}f. In this buoyancy-dominated limit, the natural convection dominates the shadowgraphy, and the natural flow rates far exceeds the forced convection. Instead, the flow much resembles purely natural convection. Figure~\ref{fig: dispbh_U} shows the reconstruction of the $x$-direction velocity for cases R1:n - R1:bh4. Initially, a near-zero magnitude flow velocity is reconstructed for neutrally buoyant case R1:n in figure~\ref{fig: dispbh_U}a. Here, the mean downstream flow imposed by the natural circulation loop is not captured by the shadow image velocimetry as it lacks contrast due to the absence of any fluctuations in intensity. As heating is applied the bottom surface, the channel flow is characterised by increasingly strong secondary flows. At first, for case R1:bh1, moderate counter flow can be observed near the heated bottom surface. Eventually, for cases R1:bh2 and beyond, a positive flow rate in the downstream direction is perceived near the opposite wall. Here, thermal plumes that reach the top of the channel act effectively as tracers of fluid motion, yielding a non-zero displacement field throughout the domain. As can be seen in figures \ref{fig: dispbh_U}c-e, the two near-wall regions show velocities of opposite signs. Here, the significant dilation of the carbon dioxide has led to the formation of a cell in which the fluid recirculates continuously, with a cell width much larger than the current field of view. 

\begin{figure}
    \centering
    \begin{subfigure}[t]{0.235\textwidth}
         \centering
         \includegraphics[width=\textwidth]{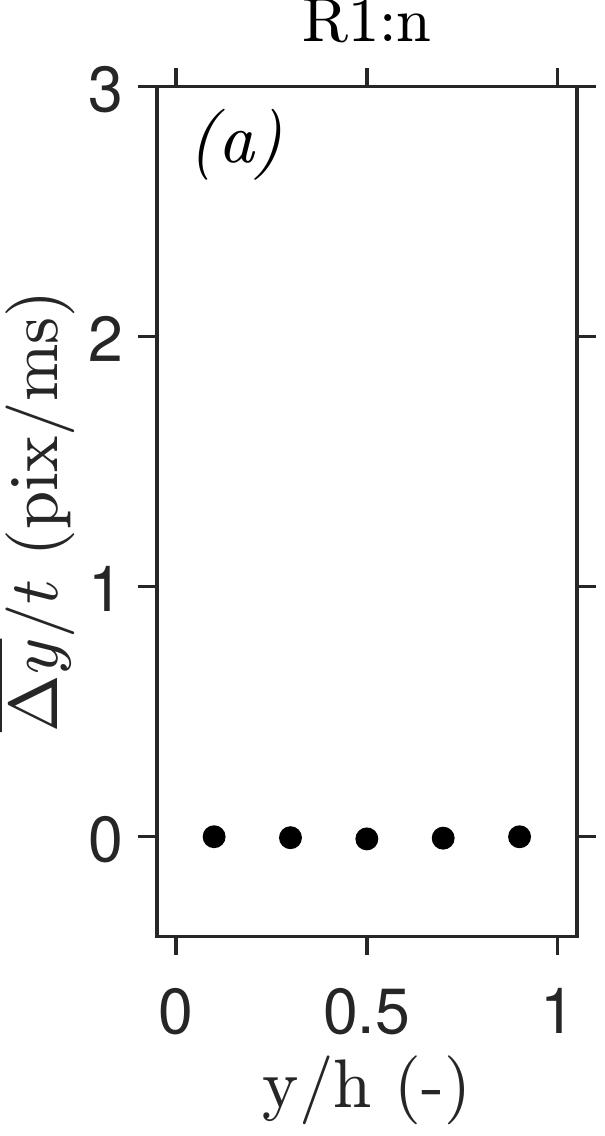}
         %\label{fig: dispbh1}
    \end{subfigure}
    \hfill
        \begin{subfigure}[t]{0.18\textwidth}
         \centering
         \includegraphics[width=\textwidth]{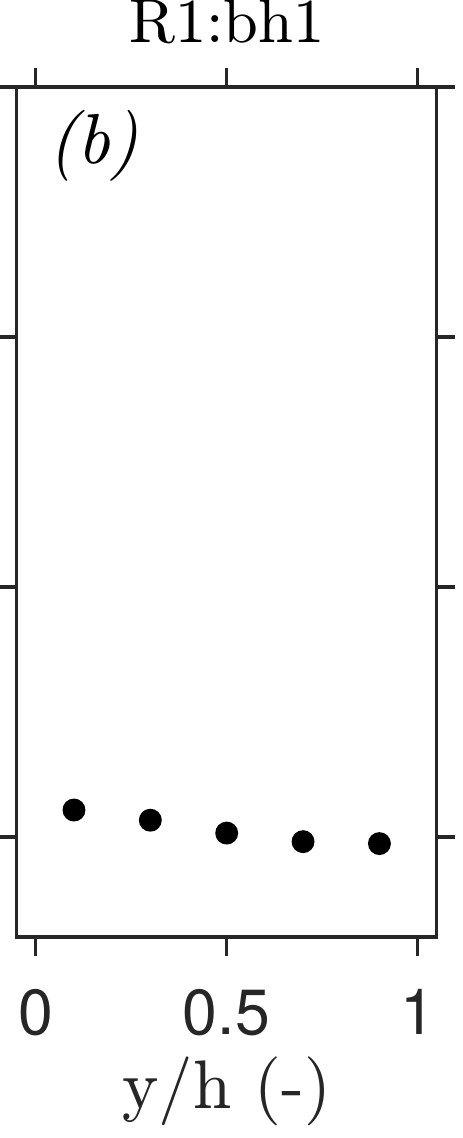}
         %\label{fig: dispbh2}
    \end{subfigure}
    \hfill
            \begin{subfigure}[t]{0.18\textwidth}
         \centering
         \includegraphics[width=\textwidth]{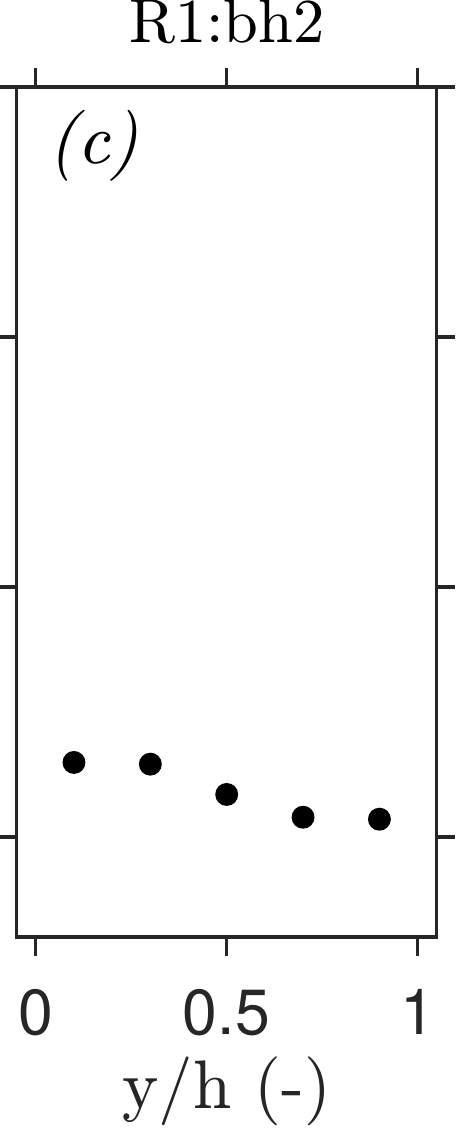}
         %\label{fig: dispbh3}
    \end{subfigure}
    \hfill
            \begin{subfigure}[t]{0.18\textwidth}
         \centering
         \includegraphics[width=\textwidth]{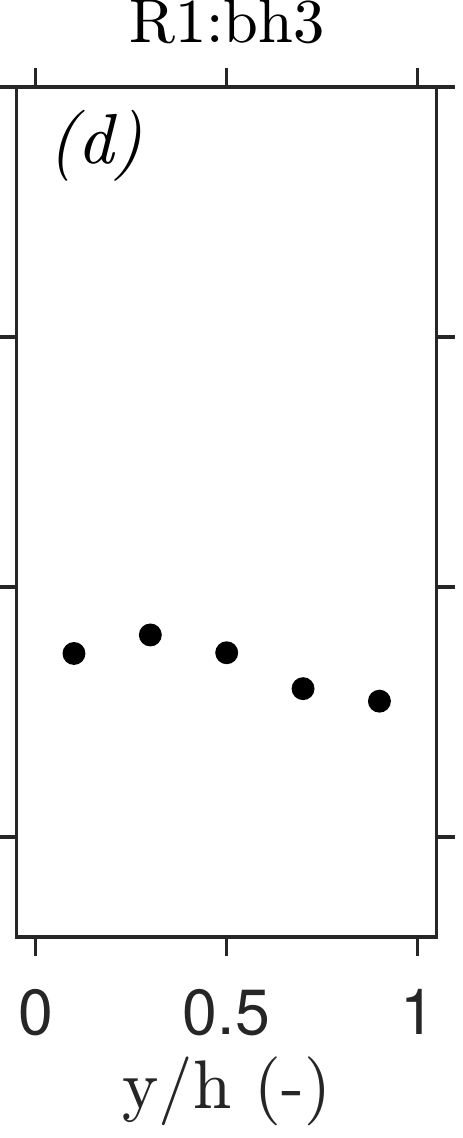}
         %\label{fig: dispbh4}
    \end{subfigure}
    \hfill
            \begin{subfigure}[t]{0.18\textwidth}
         \centering
         \includegraphics[width=\textwidth]{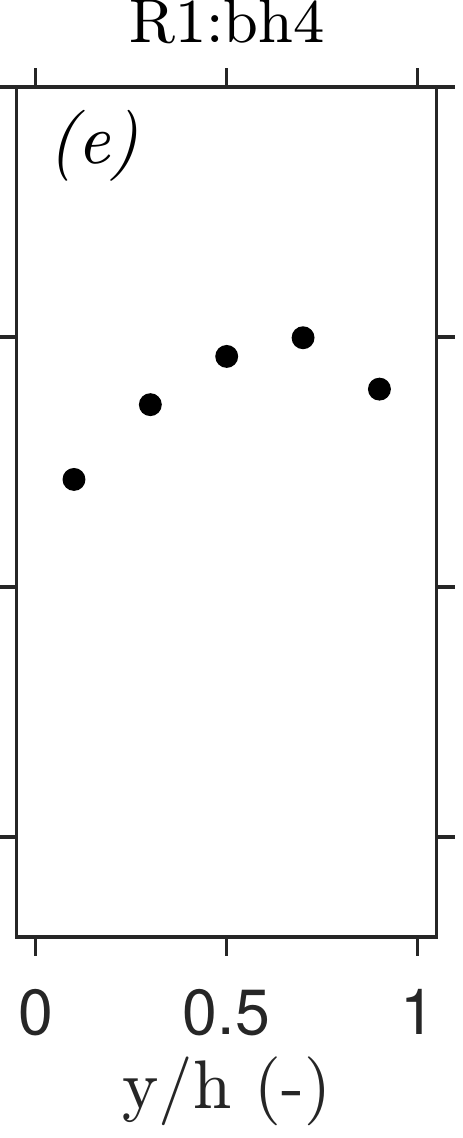}
         %\label{fig: dispbh5}
    \end{subfigure}
    \caption{Mean perceived vertical displacements $\overline{\Delta y}/t$ for case $R1$, with progressively increased bottom-up heating rates in the consecutive subfigures. 
    % The data is shown with errorbars of $\pm 2\sigma/\sqrt{N}$, where $N$ is the number of frame pairs.
    }
    \label{fig: dispbh}
\end{figure}

In the unstably stratified configuration, the velocity of the upward plumes depends on the applied heating rate. The mean displacement rates of the plume motion is shown in figure~\ref{fig: dispbh}. For cases R1:bh1 and R1:bh2, moderate and positive non-zero motion is only observed in the vicinity of the heated wall, in which compressible plumes are perceived in the shadowgrams. The vertical velocity of the plumes increases as the heating rate is increased from R1:bh1 to R1:bh2. As the heating rate is further incremented to case R1:bh3, finite upward motion is seen throughout the channel. Here, the thermal plumes traverse the test section channel at a larger velocity than for cases R1:bh1 and R1:bh2. As the amount of plumes that is produced increases further, the uncertainty in the deduction of their mean velocity also increases. Eventually, for case R1:bh4, the upward velocity is equal in magnitude to the maximum $x$-direction velocity, as can be deducted from a comparison between figures \ref{fig: dispbh_U}e and \ref{fig: dispbh}e. Importantly, the results of figure \ref{fig: dispbh} do not necessarily imply that a net upward flow is present in the channel. Instead, figure \ref{fig: dispbh} shows the motion of refractive structures somewhere within the channel. 

\subsection{Stable stratification}\label{sec:res:stable}
\subsubsection{Turbulence attenuation}
\begin{figure}
    \centering
    \begin{subfigure}[t]{0.47\textwidth}
         \centering
         \includegraphics[width=\textwidth]{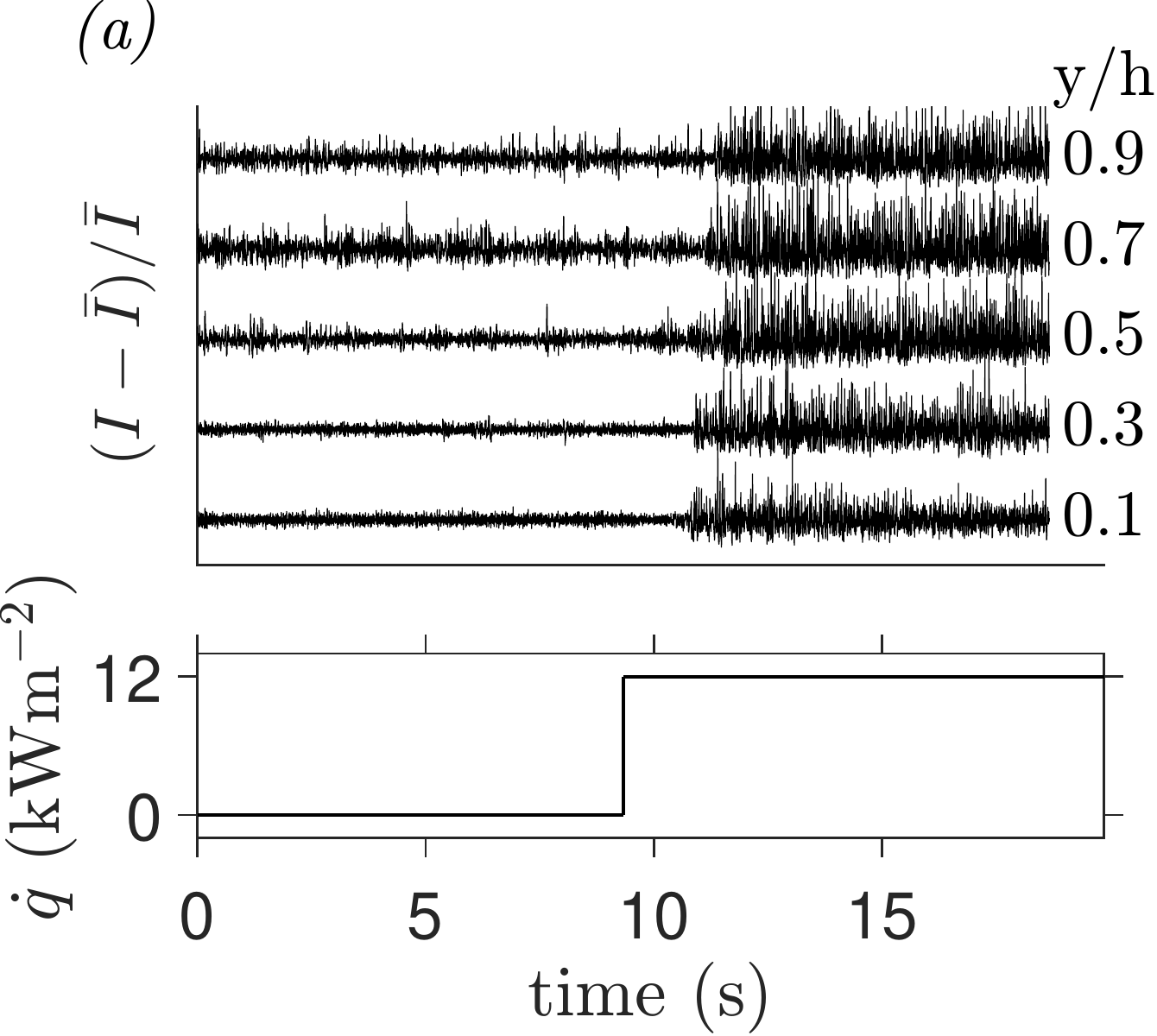}
         %\label{fig: fluctimebot}
    \end{subfigure}
        \begin{subfigure}[t]{0.47\textwidth}
         \centering
         \includegraphics[width=\textwidth]{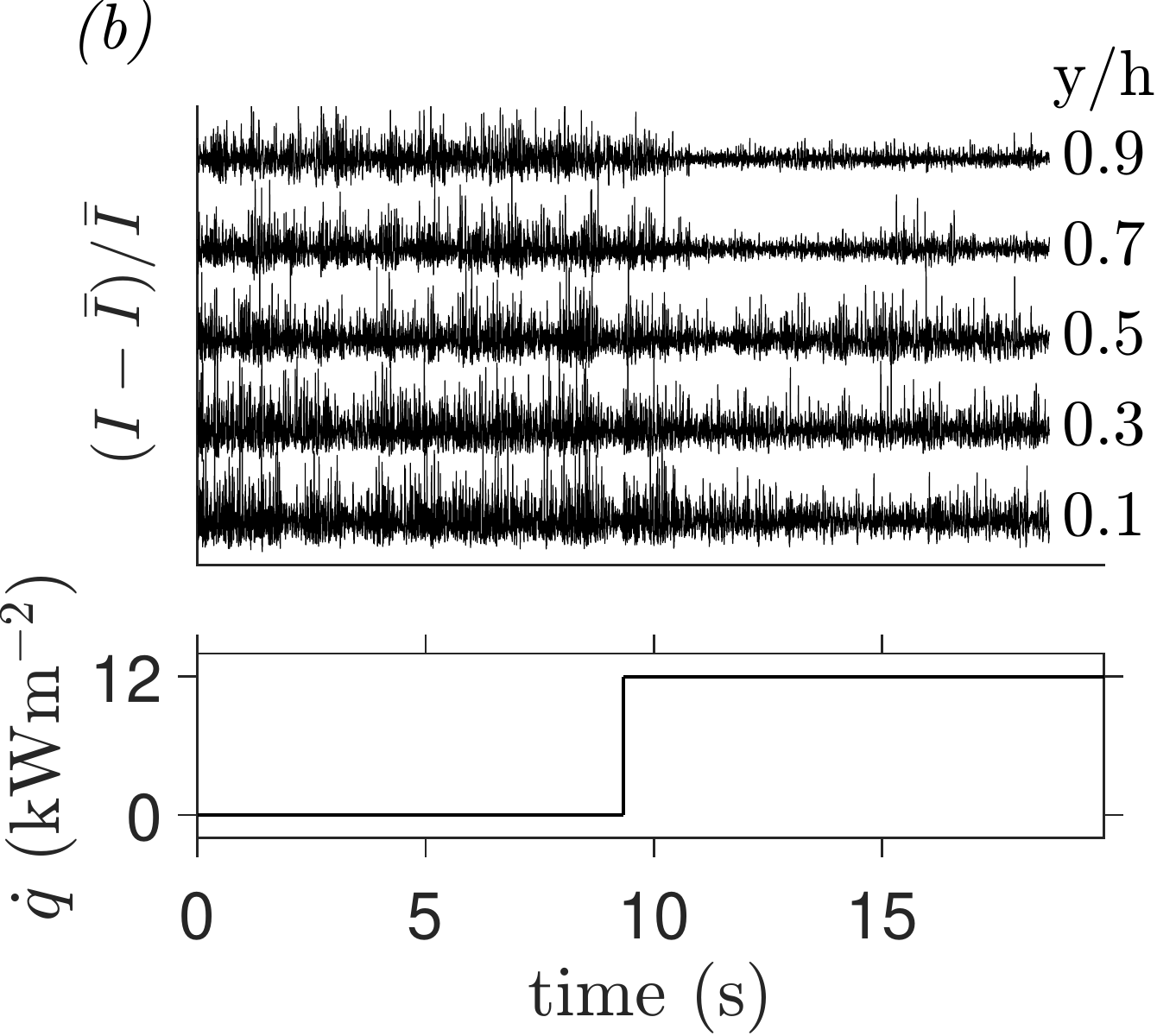}
         %\label{fig: fluctimetop}
    \end{subfigure}
    \caption{Image brightness $I$ over time. In subfigure (a) the carbon dioxide is heated from the bottom upwards from an initial state R2:n. In subfigure (b) the carbon dioxide is heated from the top downwards from an initial state R5:n.}
    \label{fig: timefluct}
\end{figure}

A process in which fluctuations and vertical movement are diminished follows from the top-down heating of the highly property-variant fluid for the cases in figure~\ref{fig:casetop}. Therewith, the flow of carbon dioxide is modulated in a distinctly dissimilar manner than for bottom-up heating. The sharp contrast between the response of flows subject to either stratification type can be highlighted by considering several points at different vertical positions within the channel. Figure~\ref{fig: timefluct} shows the normalised shadowgram intensity of the points over time. Initially, cases R2:n and R5:n are considered in figures \ref{fig: timefluct}a and \ref{fig: timefluct}b, respectively. Here, the fluid in the channel remains unheated. Whereas the variability in the signal of case R2:n is larger in its lighter top-layer than in its denser bulk, its brightness (and therefore its density) is relatively homogeneous in time. After 9.3 seconds, a heating rate $\dot{q}$ with a magnitude of $12.0$ kWm$^{-2}$ is applied to the channel surface in the bottom-upwards configuration in figure~\ref{fig: timefluct}a. Some time after the heating has started, the unstably stratified density interface that is formed near the heated wall breaks down. A plume-laden front rises to the top of the channel follows, as is shown in figure~\ref{fig:plumes}. Eventually, the signal is saturated with frequent passages of rising plumes, evidenced by a large variability in image brightness over time.
\begin{figure}
    \centering
    \begin{subfigure}[t]{0.2093\textwidth}
         \centering
         \includegraphics[width=\textwidth]{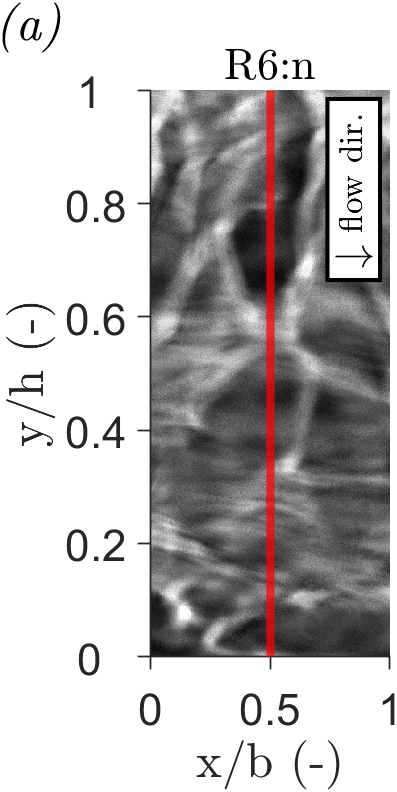}
         %\label{fig: SStopheat0}
    \end{subfigure}
    \hfill
        \begin{subfigure}[t]{0.2055\textwidth}
         \centering
         \includegraphics[width=\textwidth]{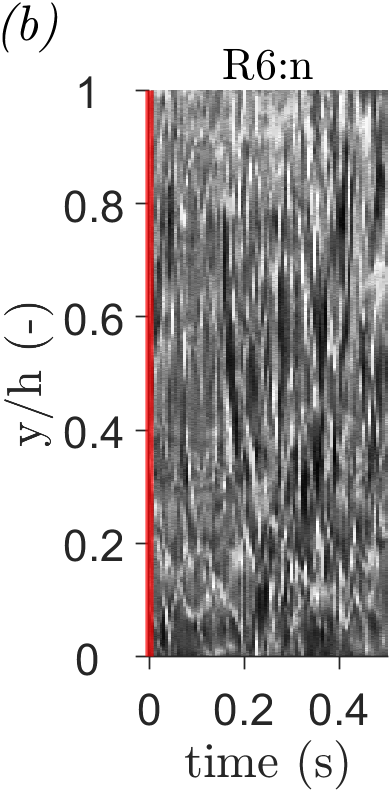}
         %\label{fig: SStopheat1}
    \end{subfigure}
    \hfill
            \begin{subfigure}[t]{0.134\textwidth}
         \centering
         \includegraphics[width=\textwidth]{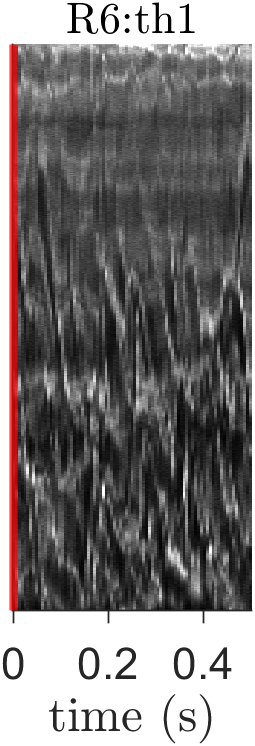}
         %\label{fig: SStopheat2}
    \end{subfigure}
    \hfill
            \begin{subfigure}[t]{0.134\textwidth}
         \centering
         \includegraphics[width=\textwidth]{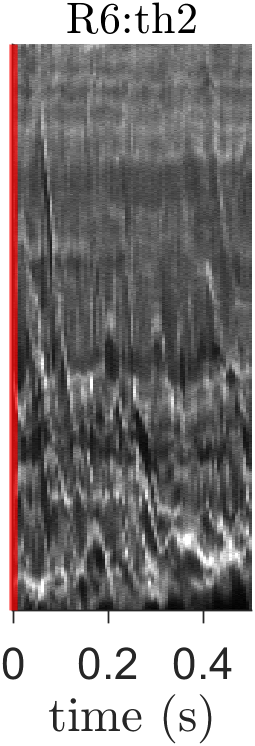}
         %\label{fig: SStopheat3}
    \end{subfigure}
    \hfill
            \begin{subfigure}[t]{0.134\textwidth}
         \centering
         \includegraphics[width=\textwidth]{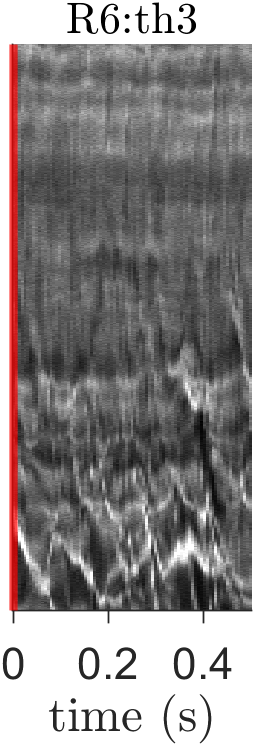}
         %\label{fig: SStopheat4}
    \end{subfigure}
    \hfill
            \begin{subfigure}[t]{0.134\textwidth}
         \centering
         \includegraphics[width=\textwidth]{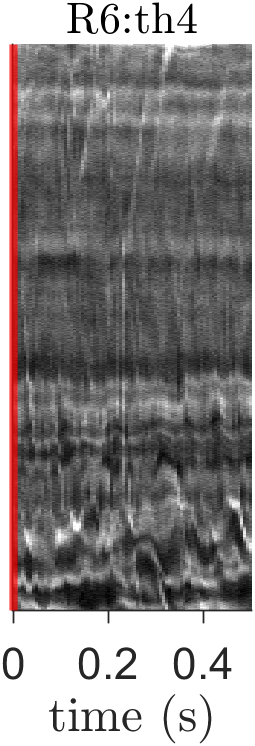}
         %\label{fig: SStopheat5}
    \end{subfigure}
    \hfill
    \caption{Space-time (b) representation of the optical signal at the vertical line that is indicated in the instantaneous shadowgram shown in (a). The top-down heating rate is progressively increased for the consecutive cases shown in (b).}
    \label{fig: SStopheat}
\end{figure}

Shadowgrams of case R5:n, in which a larger mass flow rate is applied through the test section, are characterized by the intermittent passage of shadowgraph structures in an unheated setting. As the passage of refractive structures corresponds with consecutive maxima and minima in image brightness, the variability of the signal far exceeds that of case R2:n. Similarly, a heating rate $\dot{q}$ with a magnitude of ${12.0 \text{ kWm}^{-2}}$ is applied to the channel surface in the top-down configuration after 9.3 seconds. Consequently, the variation in shadowgraphy signal in the region near the heated wall decreases significantly. Here, a region that is mostly homogeneous in brightness (and hence density) is formed. The layer is characterised by a near-complete suppression of fluctuations, with the occasional passage of a single refractive streak at moderate heating rates. The emergence of a homogeneous layer can also be observed in figure~\ref{fig: SStopheat}b, in which the signal along the red, vertical line in figure~\ref{fig: SStopheat}a is shown over time. Here, the top-down heating rate is incrementally increased from R6:n to R6:th4. As the heating rate is increased, the non-fluctuating region expands away from the heated wall to cover a greater fraction of the channel height. As shown in figure~\ref{fig:casetop}, the thickness of the layer furthermore depends on the imposed Reynolds number. Here, the homogeneous layer becomes more limited in size as the Reynolds number is increased.

When the top surface of the flow is heated, the shape of the refractive structures in the shadowgrams changes. The leftmost column in figure~\ref{fig:casetop} shows the channel flow of $\text{CO}_2$ in an unheated state. Streaks of maxima in shadowgram brightness outline the boundaries of homogeneous refractive structures. When unheated, the flow contains structures of any shape and orientation, exhibiting gradients in density in any direction. After heating is applied, the imaged streaks are found to re-orient within the channel. Structures that represent a constant density that initially span a finite vertical height, are increasingly stretched in the streamwise direction. As a result, the mean angle of the streaks with the horizontal axis reduces. As such, the fraction of the height of a channel spanned by a structure of constant density decreases. In the near-wall layer, vertically oriented streaks disappear altogether, revealing an axially homogeneous density field. Here, the originally irregular density field is assumed to be redistributed in the channel as the flow is heated.
\begin{figure}
    \centering
    \begin{subfigure}[t]{0.675\textwidth}
         \centering
         \includegraphics[width=\textwidth]{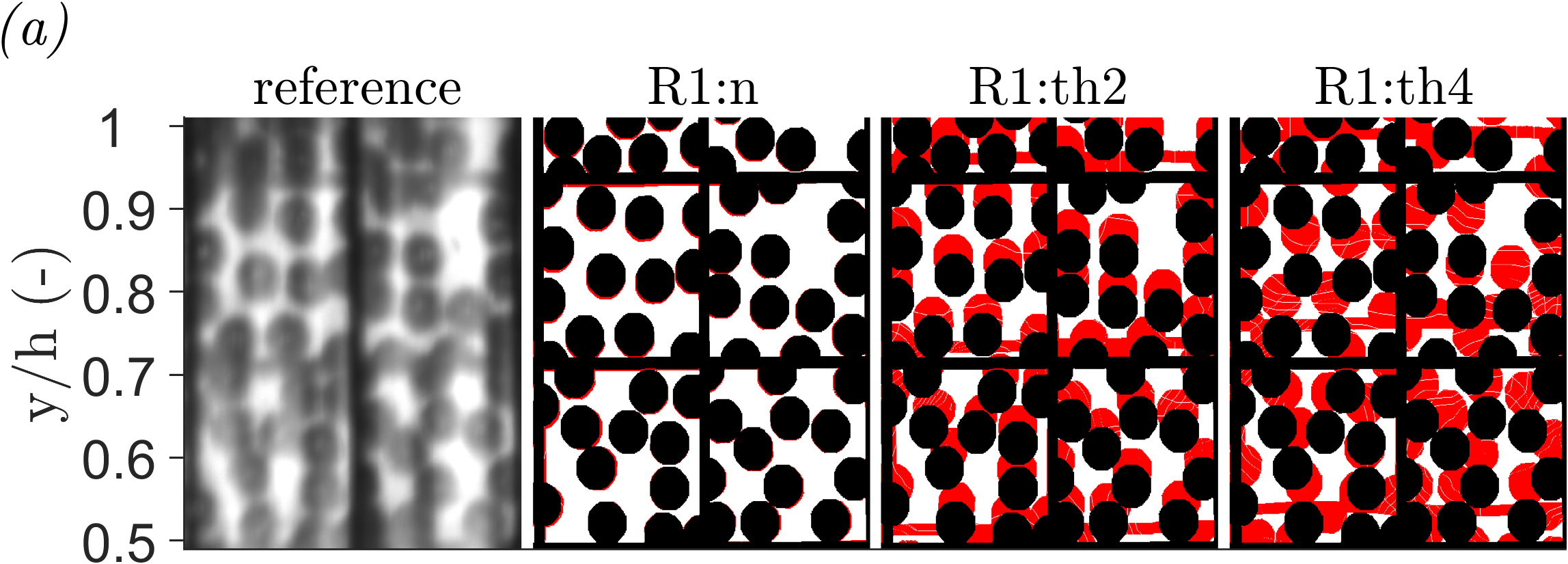}
         %\label{fig: BOSimg}
    \end{subfigure}
    \hfill
        \begin{subfigure}[t]{0.3\textwidth}
         \centering
         \includegraphics[width=\textwidth]{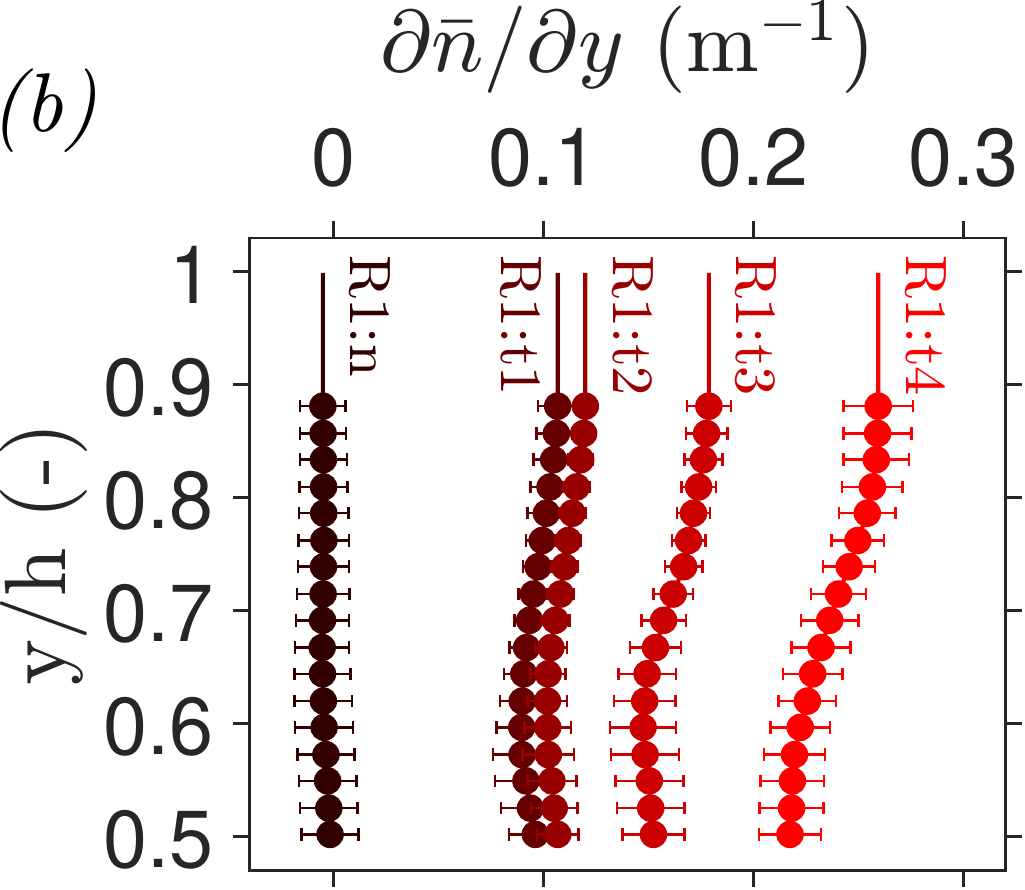}
         %\label{fig: BOSlines}
    \end{subfigure}
    \caption{The displaced state of an in-focus pattern that is imaged through carbon dioxide at supercritical pressure is shown for various top-down heating rates for case $R1$ in (a). The original pattern is shown in black. The mean, interpolated local displacement for each point is used to shift the black pattern, yielding the pattern shown in red. Profiles of the path-integrated vertical gradients in refractive index are shown in (b). The data in (b) is shown with errorbars of $\pm 2\sigma$.}
    \label{fig: BOS}
\end{figure}

\begin{figure}
    \centering
    \begin{subfigure}[t]{0.238\textwidth}
         \centering
         \includegraphics[width=\textwidth]{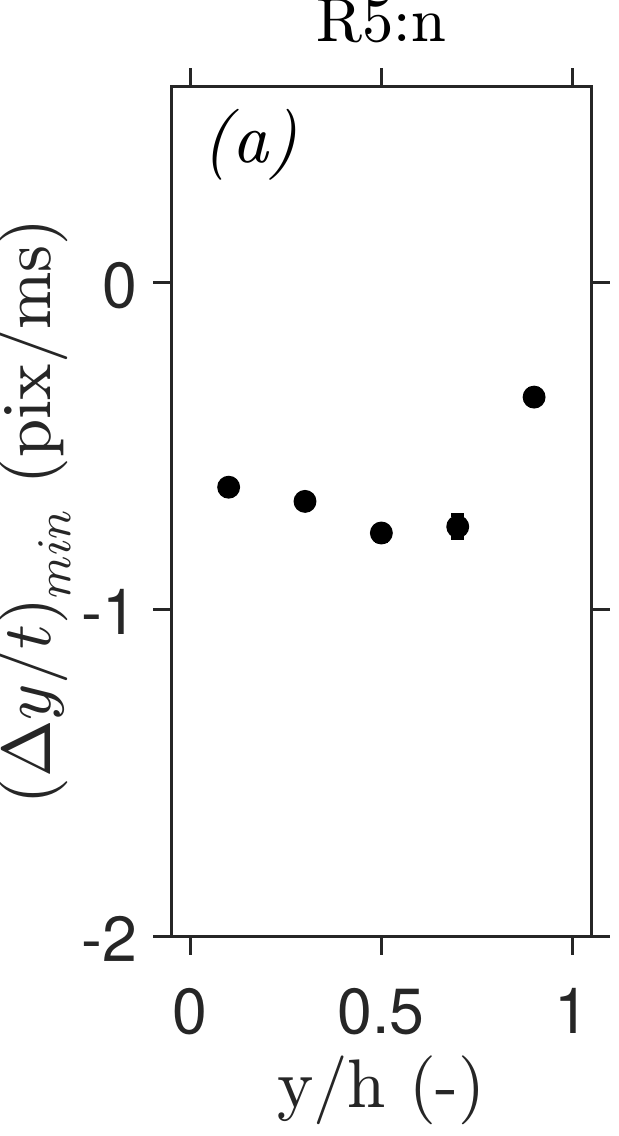}
         %\label{fig: dispth1}
    \end{subfigure}
    \hfill
        \begin{subfigure}[t]{0.178\textwidth}
         \centering
         \includegraphics[width=\textwidth]{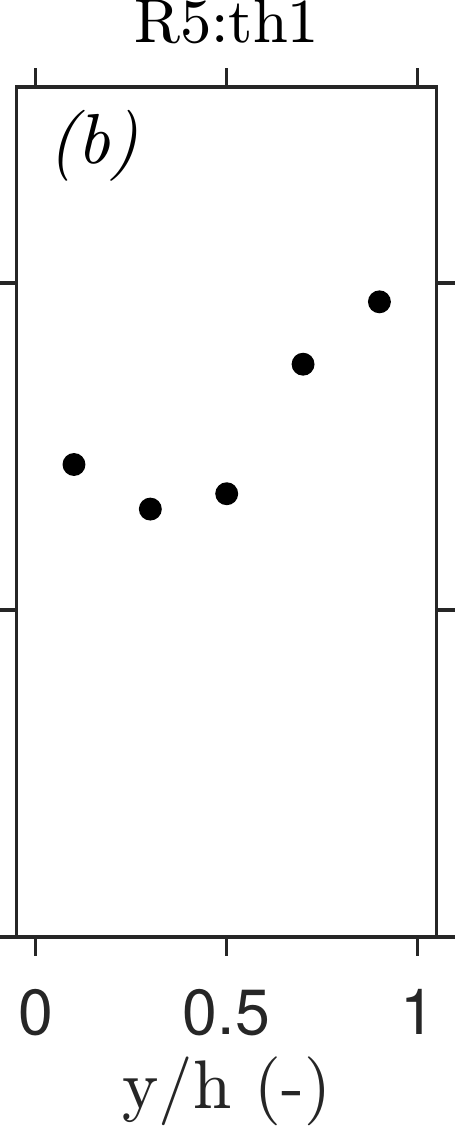}
         %\label{fig: dispth2}
    \end{subfigure}
    \hfill
            \begin{subfigure}[t]{0.178\textwidth}
         \centering
         \includegraphics[width=\textwidth]{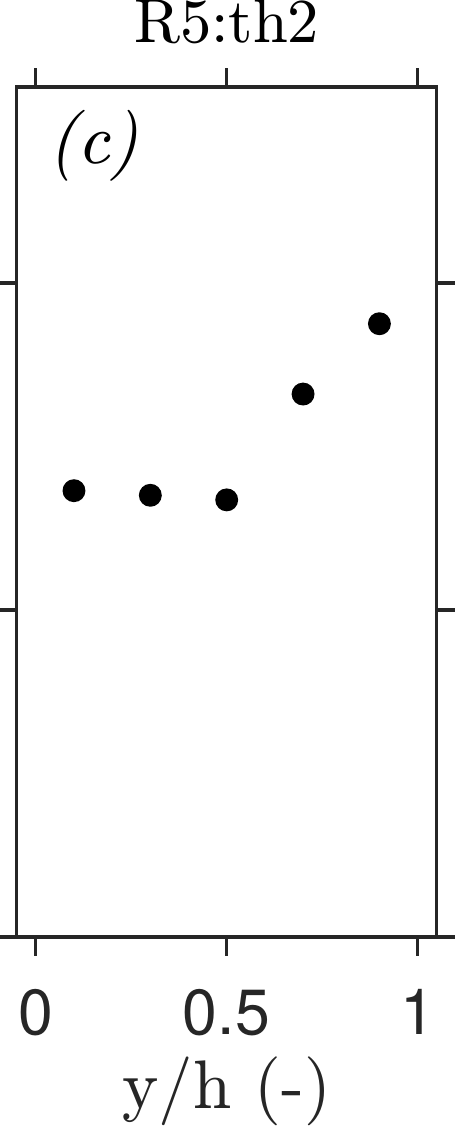}
         %\label{fig: dispth3}
    \end{subfigure}
    \hfill
            \begin{subfigure}[t]{0.178\textwidth}
         \centering
         \includegraphics[width=\textwidth]{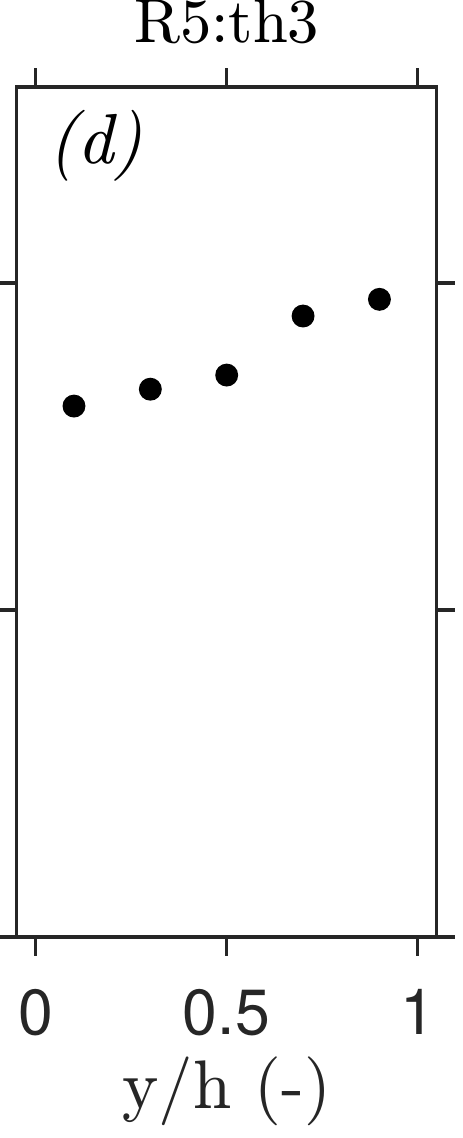}
         %\label{fig: dispth4}
    \end{subfigure}
    \hfill
            \begin{subfigure}[t]{0.178\textwidth}
         \centering
         \includegraphics[width=\textwidth]{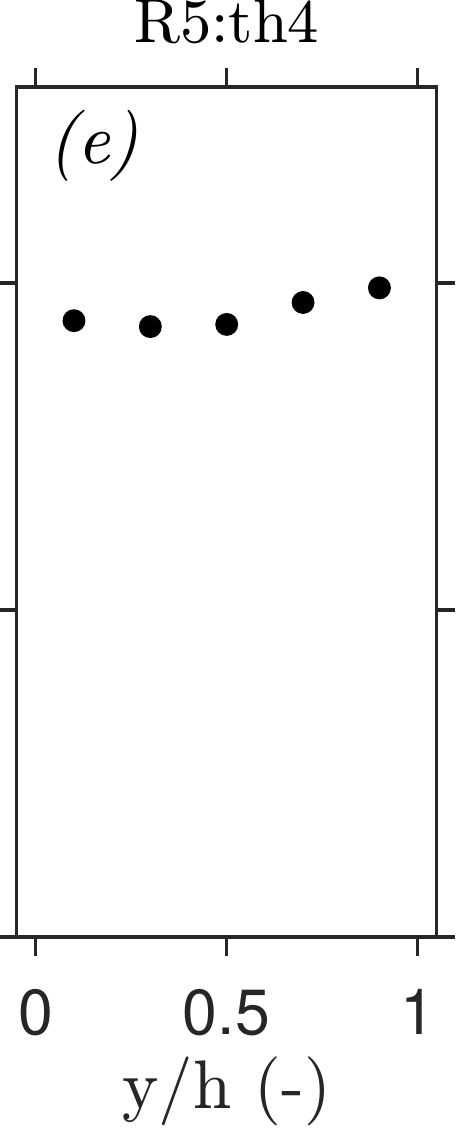}
         %\label{fig: dispth5}
    \end{subfigure}
    \caption{Maximum instantaneous vertical displacements in the negative $y$-direction for case $R5$, with progressively increased top-down heating rates in the consecutive subfigures. 
    % The data is shown with errorbars of $\pm 2\sigma/\sqrt{N}$, where $N$ is the number of frame pairs.
    }
    \label{fig: dispth}
\end{figure}

Figure~\ref{fig: BOS} reveals that the initial density field is redistributed along a mean density gradient, imposed by the top-down heating. For the purpose of highlighting such mean gradients in refractive index, the BOS configuration indicated in cross section~\textbf{B} of figure~\ref{fig:schematic} is considered. In figure~\ref{fig: BOS}, the deformation of a background pattern by the refractive $\text{CO}_2$ is considered. Only under stable stratification do the fluctuations in the optical signal attenuate sufficiently to get a clear BOS image. The considered reference pattern, shown in black in figure~\ref{fig: BOS}a, is printed onto a transparent sheet and placed behind the $\text{CO}_2$. As the carbon dioxide is heated, the pattern's image deforms. The deformed images for cases R1:n, R1:th2 and R1:th4 are shown in red in figure~\ref{fig: BOS}a. Here, the black pattern is shifted with the correlated deformation displacements at any point, yielding the deformed pattern shown in red. In neutrally buoyant case R1:n, the imaged pattern remains in a predominantly undeformed state. As heating is applied to the top of the channel in cases R1:th2 and R1:th4, the dot pattern in the top half of the channel is stretched by variations in the refractive index. Here, the background pattern is deformed to a greater extent when the applied heating rate is increased. Figure~\ref{fig: BOS}b shows the vertical gradients in path-integrated refractive index that are deducted from the dot pattern displacements. Here, $n_0=1.17, M=0.5$, and $Z_{\text{D}}=8.9\cdot 10^{-2}$ m. The magnitude of the refractive index gradients are found to increase as greater heating rates are applied to the channel flow. Furthermore, the gradients are the largest in the near-wall region for all considered cases. The presence of a strong vertical gradient in $\bar{n}$ suggests the existence of an equivalent gradient in $\text{CO}_2$ density $\bar{\rho}$ in the current top-down heated configuration. Given the positive values of $\partial \bar{n}/\partial y$ in figure~\ref{fig: BOS}b, the density of the carbon dioxide increases away from the heated wall. The imposed vertical gradient in $\bar{\rho}$ imposes a buoyant force on the fluid in the channel. As discussed by \citet{garcia-villalba2011}, fluid parcels of a certain density are pushed towards a preferential vertical position along the constant gradient. If the buoyant force exceeds the inertial forces of a fluid parcel, its vertical movement is diminished once it is at its preferential position. This is true for an increasingly large density range and for an increasingly large portion of the channel as the imposed density gradient becomes sharper when the heating rate is increased. As a result, vertical motion is increasingly diminished in the near-wall region when heating is applied, as also shown in figure~\ref{fig: dispth}. In the figure, the largest instantaneous downward shadow displacements within the considered measurement interval are shown for cases R5:n - R5:th4. Initially, for unheated case R5:n, the flow is characterised by the sporadic vertical movement of compressible eddies. As moderate heating is applied in cases R5:th1 and R5:th2, vertical motion in the near-wall region is suppressed along the vertical density gradient. The region within which the vertical motion of turbulent eddies is suppressed grows as the heating rate is increased towards case R5:th3. Eventually, most vertical motion is suppressed in case R5:th4. 

\subsubsection{Implications on PIV}
The large deformation of the background pattern reveals limitations to the further characterization of the flow at a near-pseudo-critical state beyond the spatial resolution that the current thermal tracers allow for. In particular, the large image distortions affect the applicability of Particle Image Velocimetry (PIV), where a correlation of the displacements of tracer particles is used to re-construct a flow field. The images of carbon dioxide at the current near-pseudo-critical conditions are characterised by large deformations both when the $\text{CO}_2$ is heated locally, and when refractive structures that correlate with the fluid motion are present in an unheated turbulent flow. Furthermore, the gradients of these deformations can be particularly large for heated turbulent flows, leading to image blur. Additionally, total internal reflections across stratified layers can lead to the local compression of the object in the image plane, obscuring its features. As a result, the apparent displacements of the particles may not correspond to their actual displacements. This is most prevalent as the tracers move across density gradients within the flow. The current back-projected error in determining the particle positions within the channel can be of the order of 100 $\mu m$. Therewith, it can exceed the particle diameters of the tracers \citep{valori2019,adamheadPIV} that are relatively small to reduce slip velocities with the property-variant working fluid. As such, PIV systems may prove to be particularly difficult to calibrate when a flow displays large local variations in thermodynamic properties.
\subsection{Overall heat transfer}\label{sec:res:heat}
\subsubsection{Surface temperature data}
\begin{figure}
    \centering
\includegraphics[width=\textwidth]{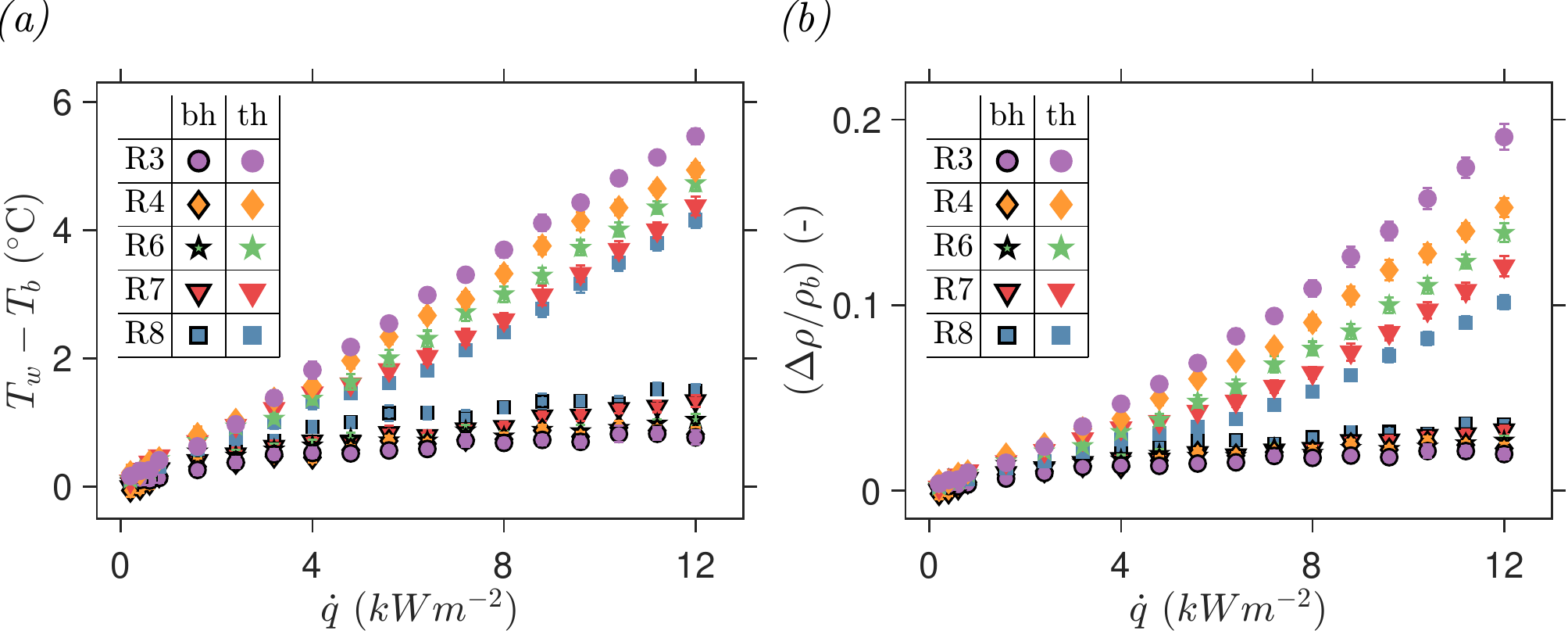}
\caption{Wall temperature (a) and deducted wall density data (b) for a selection of cases with different nominal Reynolds numbers from figures \ref{fig:casebot} and \ref{fig:casetop}. Bottom-up (bh) heating data points are shown with black-outlined markers, whereas top-down (th) heating data points are shown with homogenous markers. The data is shown with errorbars of $\pm 2\sigma$.}
\label{fig:dT drho}
\end{figure}

Measurements of the surface temperature of the heated walls serve to provide a quantitative context to the optical results presented in this work. Figure~\ref{fig:dT drho}a shows the mean increase in wall temperature $T_{\text{w}}$ with respect to the constant bulk temperature $T_{\text{b}}$ for a selection of cases of either configuration. When the $\text{CO}_2$ is unstably stratified, the increase in surface temperature remains within 1.5$^{\circ} C$ for the shown parameter range. Furthermore, the measured step in wall temperature decreases as the imposed heat flux is increased. Therefore, the heat transfer rate is expected to progressively improve as greater heating rates are imposed. Initially, a qualitatively similar trend is observed when the top surface of the channel is heated. However, beyond a threshold heat flux, a near-linear increase in wall temperatures (indicative of a predominantly constant heat transfer coefficient) is found for increasing heating rates. This threshold heat flux decreases with decreasing Reynolds numbers. Beyond ${\dot{q}=4\text{ kWm}^{-2}}$, the near-wall temperatures of the two heating configurations start to significantly differ for any of the considered Reynolds numbers. Eventually, for the largest heating rate considered in this work, the increase in surface temperature of the stable stratification exceeds the unstable stratification's temperature increase by twofold. As such, heat is removed much more effectively from a heated wall in an unstably stratified setting than it is in a stably stratified configuration for the current experimental conditions.

Figure~\ref{fig:dT drho}b shows the relative expansion of the carbon dioxide in the vicinity of the heated wall, as deducted from the wall temperature data of figure~\ref{fig:dT drho}a. In accordance with the wall temperature data, the heating-induced density change is significantly larger for stable stratification than it is for unstable stratification. As such, the degree of stratification (expressed by the Richardson number) is greater when the stratification is stable than when it is unstable under equivalent heating, as previously noted in figure~\ref{fig: Ri-Re}. The relative difference in dilation even exceeds the relative difference in surface temperature increase, as can be deducted from the increasingly superlinear trend in density variation during stable stratification. Here, the carbon dioxide exhibits increasingly large variations in density as it is heated towards the non-linear pseudo-critical curve from the liquid-like test section inlet conditions. 
\subsubsection{Characterization of heat transfer}
\begin{figure}
    \centering
\includegraphics[width=\textwidth]{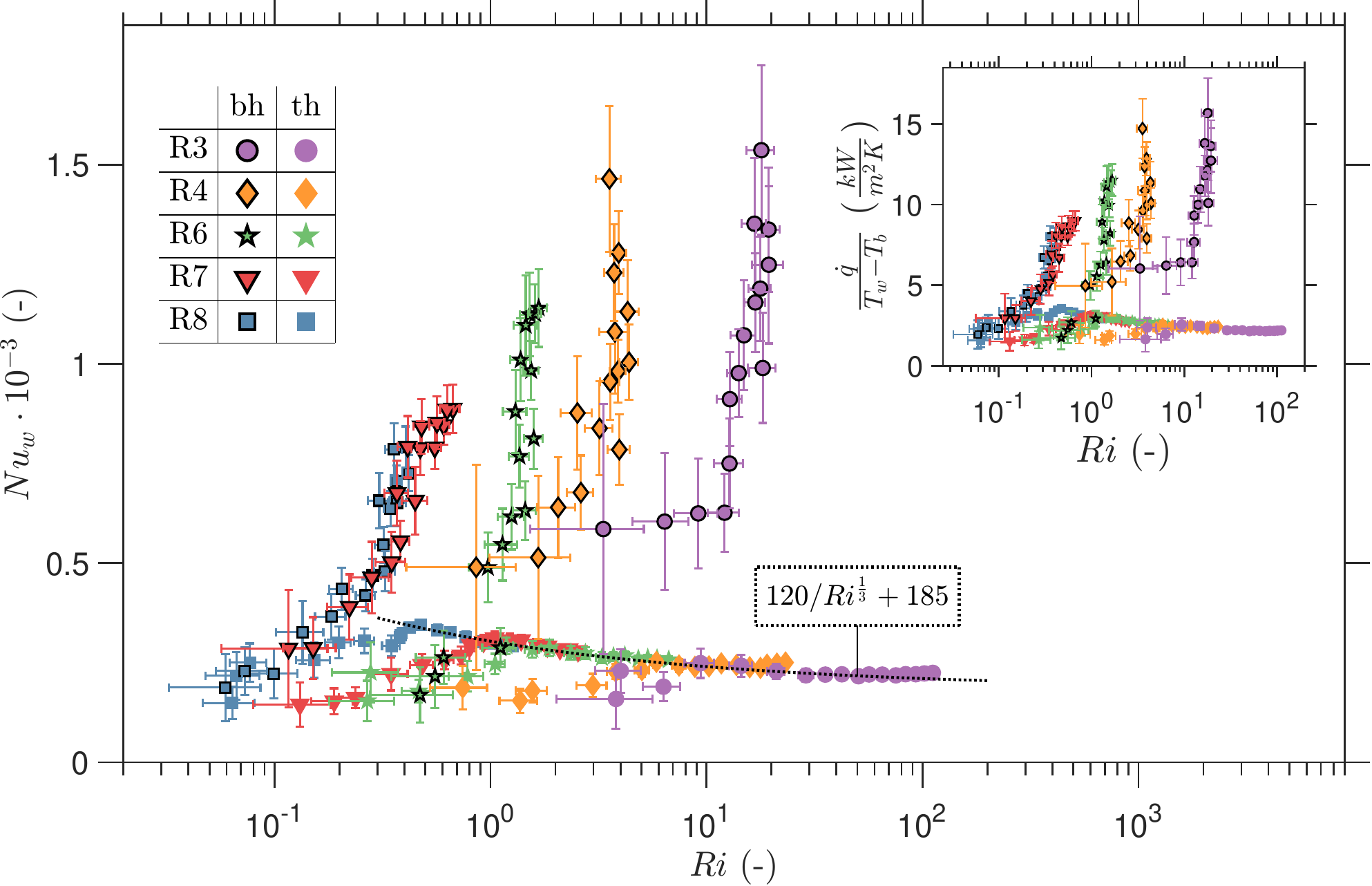}
\caption{Nusselt number (main figure) and heat transfer coefficient data (top right corner) for a selection of cases from figures \ref{fig:casebot} and \ref{fig:casetop}, as a function of the deducted value for $Ri$. Bottom-up (bh) heating data points are shown with black-outlined markers, whereas top-down (th) heating data points are shown with homogenous markers. The data is shown with errorbars of $\pm 2\sigma$.}
\label{fig: h Ri}
\end{figure}

An analysis of the empirical heat transfer reveals trends that are in qualitative agreement with the optical data presented in this work. Figure~\ref{fig: h Ri} shows the mean Nusselt number $Nu_{\text{w}}$ as a function of the Richardson number $Ri$ for a selection of cases of either configuration, with
\begin{equation}
    Nu_{\text{w}}=\frac{\dot{q}}{T_{\text{w}}-T_{\text{b}}}\cdot\frac{h}{\lambda_{\text{w}}}.
\end{equation}
Here, $\lambda_{\text{w}}$ is the near-wall thermal conductivity, which is evaluated at $T_{\text{w}}$ and $P_{\text{b}1}$. The top-right corner of the figure furthermore shows the evolution of the measured heat transfer coefficient with $Ri$. However, as the variation in $\lambda_{\text{w}}$ is moderate within the current parameter range, the dependence of both the heat transfer coefficient and $Nu_{\text{w}}$ on $Ri$ are similar. When buoyancy is non-dominant, at $Ri\lesssim 0.1$, the empirical heat transfer coefficients of the two stratification types are of a similar magnitude. However, as buoyancy becomes non-negligible for larger heating rates, the values for $Nu_{\text{w}}$ diverge. 

When the carbon dioxide is heated from the bottom upwards, $Nu_{\text{w}}$ improves with respect to its neutrally buoyant value. The shadowgrams of unstably stratified $\text{CO}_2$ (shown in figure \ref{fig:casebot}) are characterized by the strong secondary flows that are elaborated on in figures \ref{fig: SSbotheat}-\ref{fig: dispbh}. In the limit of small Reynolds numbers, plume motion is observed to form recirculation cells that locally counteract the imposed forced convection. For any $Re_{Dh}$, the expansion of $\text{CO}_2$ causes motion in the wall-normal direction, in which parcels of lighter fluid are locally lifted towards a denser bulk flow. Here, hotter fluid is continuously carried away from the heated wall, replacing it with a denser and colder fluid that is recirculated. Therewith, the mean near-wall fluid temperature decreases with respect to an equivalent neutrally buoyant setting, leading to the sub-linear increase of ${(T_{\text{w}}-T_{\text{b}})}$ in figure~\ref{fig:dT drho}a. Furthermore, the vertical plume motion increases the wall-normal momentum transport in the channel. Consequently, the vertical energy transport also increases with respect to purely forced convection, as characterized by the enhancement of $Nu_{w}$ in figure \ref{fig: h Ri}. However, this effect diminishes as the relative inertial contributions increase. Therefore, $Nu_{\text{w}}$ decreases with increasing $Re_{\text{Dh}}$ at a set heating rate. Contrary to a previous empirical investigation of purely bottom-upwards heated micro-channels of $\text{CO}_2$ with $Ri\lessapprox 1$ by \citet{whitaker2024flow}, the relative enhancement of the heat transfer progressively increases as buoyant forces become increasingly dominant at larger $Ri$ within the current parameter range. A progressive improvement of the heat transfer with increasing degrees of unstable stratification is however in accordance with the current optical data.  Here, the spatial density of the plumes and their lift velocities are found to both increase at greater heating rates, increasing the vertical momentum transfer in the channel. As such, the progressive enhancement of the heat transfer as buoyancy becomes more dominant with increasing degrees of unstable stratification in figure~\ref{fig: h Ri} is expected from the increasingly dominant motion of the thermal plumes.

As the carbon dioxide is heated in the top-down orientation in figure \ref{fig:casetop}, the heat transfer is not enhanced beyond an optimum. Instead, when $Ri$ is increased beyond a threshold value, the heat transfer deteriorates. In the shadowgrams, the imposed density gradient of figure \ref{fig: BOS} is found to counteract vertical movement as the carbon dioxide is heated at the top surface of the channel, as is elaborated in figures \ref{fig: SStopheat} and \ref{fig: dispth}. The resulting flow is characterized by the suppression of intermittent, irregular motion. Here, the mixing of distinct layers with different densities is counteracted. As such, the lighter fluid that is formed at the heated top wall remains predominantly in the near-wall region. As a result, the overall near-wall temperature increases. Therewith, the wall-normal heat transfer rate deteriorates in comparison with a neutrally buoyant flow, in which the wall-normal movement of warmer eddies removes hot fluid from the heated wall. Notably, at large values of $Ri$, the measured values of $Nu_{\text{w}}$ coincide in figure \ref{fig: h Ri}. Here, $Nu_{\text{w}}$ attains an asymptotic value, irrespective of the imposed Reynolds number. For these conditions, $Nu_{\text{w}}$ can be described using
\begin{equation}
    Nu_{\text{w}}=\frac{120}{Ri^{\frac{1}{3}}}+185,
\end{equation}
as indicated with the black, dashed line in figure~\ref{fig: h Ri}. As discussed above, the deterioration of heat transfer at $Ri\gtrapprox 1$ is likely a result of the suppression of fluctuations and vertical motion observed in the shadowgrams of this work. In the shadowgrams, the size of the homogenised layer grows with increasing heating rates. However, the intermittency of the optical signal remains largely constant in the near-wall region after the initial suppression of fluctuations by the thermal stratification. As such, the current moderate dependence on $Ri$ of the deterioration of the top-down heating is in agreement with the shadowgraphy. \break 

The significant non-ideality of the observed heat transfer strongly affects the performance of near-pseudo-critical heat exchangers in energy conversion systems at supercritical pressures. In heat transfer systems with ideal fluids, the buoyant forces that can be induced by the expansion of an ideal medium are generally of negligible magnitude compared to the inertial forces within a heat exchanger. As the dilation of an ideal fluid is a weak function of temperature, a buoyancy dominated flow is seldomly reached in any practical setting. Similarly, the data in figure~\ref{fig: h Ri} would remain limited to points for which $Ri\ll 1$ if an ideal, single-phase fluid was used. As such, the heat transfer would not depend on the stratification type and strength. However, a near-pseudo-critical and therewith highly property-variant fluid is currently considered. As such, significant buoyant forces can be induced with the limited heating rates that are practically attainable. As such, the direction of the applied heat transfer strongly affects the behaviour within heat exchangers that operate in the vicinity of the pseudo-critical curve. Within the current parameter range, a significant difference between the heat transfer rates of purely top-heated and purely bottom-heated flow of carbon dioxide at supercritical pressure is observed. This difference is most notable for the larger heating rates in this work, for which the bottom-up heat transfer coefficient exceeds the top-down heat transfer coefficient by nearly a full order of magnitude, as shown in figure~\ref{fig: h Ri}. However, in the flow channels of plate heat exchangers or a Printed Circuit Heat Exchangers (PCHEs), heat is generally exchanged in multiple directions. As such, the heat transfer enhancement near the bottom wall and the heat transfer deterioration near the top wall compete when the flow is heated. If the heating rate for both surfaces is equal, much like in the current work, the degree of stratification is greater in the region near the top wall. As such, the lower-density layer near the top wall would likely not be fully disrupted by the upward movement of plumes created at the bottom wall. Therewith, the behaviour within the heat exchanger would be qualitatively similar to a pipe flow to which a constant heat flux is applied \citep{chu2016flow}. The equivalent heat transfer coefficient of the heat exchanger channel is expected to lie between the deteriorated and enhanced values of the one-sided heating configurations that are considered in this work.

\section{Conclusions}\label{sec:conclusions}
In this work, the modulation of the turbulence and the stability of a continuous flow of a supercritical fluid by non-negligible buoyancy is investigated. Such flows are for instance prevalent in novel energy conversion systems that operate at supercritical pressures, which have gained particular recent interest for their potential to sustainably produce power and industrial heat. Within the heat exchangers of these energy systems in particular, non-negligible buoyancy may lead to highly non-ideal and configuration dependent heat transfer.

The current experimental facility employs the natural circulation of carbon dioxide that is passed through an optically accessible test section. To study the hydrodynamically developed flows of carbon dioxide at supercritical pressures within the test section, shadowgraphy is used in parallel with surface temperature measurements. The shadowgraphy is used to visualise a side-view, path integrated flow field. When the temperature of the $\text{CO}_2$ is sufficiently heterogeneous, compressible structures within the flow can be distinguished in the shadowgrams. As the structures are observed to move with the flow, they are considered to act as thermal flow tracers. As such, a correlation of their displacements yields velocity components in the directions perpendicular to the optical axis. It is validated that the mean streamwise displacements of these pseudo-tracers correspond to the expected mean flow velocity inside the test section, provided that the $\text{CO}_2$ is sufficiently turbulent (i.e. $Re_{\text{Dh}}\gtrsim 3\cdot10^3$) and neutrally buoyant. However, the shadow image velocimetry is characterized by relatively large uncertainties, as the method depends on the local occurrence of density fluctuations (leading to variations in refractive index) of the flow, and the finite coherence of these structures in time. Furthermore, additional, rigorous calibration is necessary to translate the shadow displacements to physical velocities when large global refractive index gradients are present in the channel under heating conditions. Nevertheless, the shadowgrams can distinguish flow movement patterns within the channel unique to either stratification type when the $\text{CO}_2$ is heated either from the bottom or from the top of the test section. As such, the shadowgraphy can serve to experimentally reveal and complement the highly transient behaviour behind the non-ideal, and thermally inert heat transfer results in existing open literature to date. 

The rectangular flow channel is heated unidirectionally in either vertical direction to impose a density gradient on the carbon dioxide within it. The $\text{CO}_2$, initially at 88.5 bar and 32.6 $^{\circ} C$, is heated towards the pseudo-critical curve. As the carbon dioxide displays sharp variations in thermodynamic properties in the vicinity of this curve, strong stratifications are observed for moderate heating rates. These strong stratifications lead to the modulation of the initially neutrally buoyant base flow, as is shown in figures \ref{fig:casebot} and \ref{fig:casetop}. Depending on the stratification type, the variability in the flow is either enhanced or suppressed. 

When the channel is heated from the bottom upwards, the flow is characterized by increasingly strong secondary movement. Here, light plumes are seen to intermittently depart from the heated surface, limiting its increase in temperature. Here, the wall-normal velocity and the spatial density of these plumes are found to increase as the applied heating rate is increased. Furthermore, as a result of the increased upward movement with increasing buoyancy, the Nusselt number is found to progressively increase with respect to the neutrally buoyant case for all considered Reynolds numbers.

On the contrary, when only the top surface of the flow channel is heated, the heat transfer deteriorates with respect to the neutrally buoyant case. This deterioration in heat transfer coincides with the suppression of oscillations in the optical signal by the stable density gradient imposed by the heating. This imposes a restriction in vertical motion and mixing across the density gradient that limits the wall-normal motion of hot $\text{CO}_2$ from the heated surface, therewith reducing its heat transfer coefficient. Notably, the heat transfer coefficient is found to no longer depend on the imposed Reynolds number beyond a threshold heating rate when the carbon dioxide is stably stratified.

The present work emphasises the necessity for the further investigation of buoyancy modulated turbulence and heat transfer in the near-pseudo-critical region beyond the current experimental conditions. Whereas the current results provide novel empirical insights in strong one-directional stratifications, the interplay of the opposing stratification remains unknown in the context of channel flows. Therefore, in order to properly describe the flow of supercritical carbon dioxide within horizontal heat exchanger channels, two-sided heating should also be considered. Furthermore, the current experimental framework can be extended to study the influence of non-negligible buoyancy in physical configurations beyond the current one. Therewith, vertical flows of highly property-variant fluids, where buoyancy also has a pronounced effect, can also be considered. 

By improving the used optical diagnostics, they can serve as references for the improvement of hydrodynamic and heat transfer models. A further increase in the spatial resolution in of the shadow velocimetry in particular can for instance be achieved with the local, \textit{in-situ} generation of thermal tracers. As an additional consequence, the flow can also be visualized when compressible tracers are not naturally present in the supercritical carbon dioxide. Furthermore, by controlling the ambient temperature of the test section, the current framework could be extended to obtain discernible optical data for a greater thermodynamic range at supercritical pressures.
\section*{Acknowledgement}
This work was funded by the European Research Council grant no. ERC-2019-CoG-864660, Critical. The authors acknowledge the assistance of M. Karsten with the electrical system, and the support of D. van Baarle with the installation of the natural circulation loop. The authors furthermore acknowledge the advice of E. Overmars on optical systems, and the fruitful discussion with B. Hoek on natural circulation.
\section*{Declaration of interests} The authors report no conflict of interest.
\appendix
\section{}\label{sec:appA}
\begin{figure}
    \centering
    \begin{subfigure}[t]{0.16\textwidth}
         \centering
         \includegraphics[width=\textwidth]{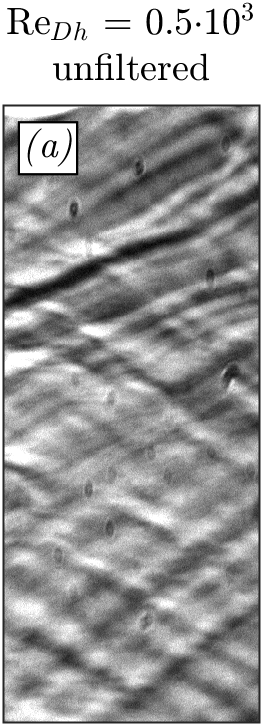}
         %\label{fig: filt1}
    \end{subfigure}
    \hfill
    \begin{subfigure}[t]{0.16\textwidth}
         \centering
         \includegraphics[width=\textwidth]{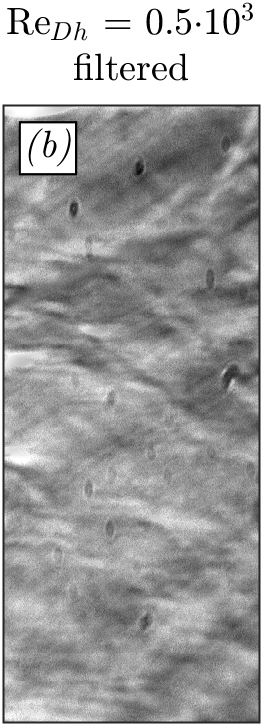}
         %\label{fig: filt2}
    \end{subfigure}
    \hfill
    \begin{subfigure}[t]{0.16\textwidth}
         \centering
         \includegraphics[width=\textwidth]{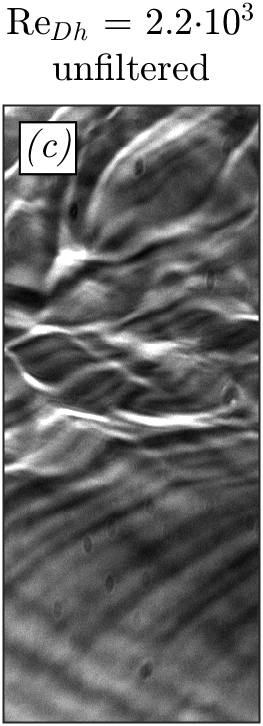}
         %\label{fig: filt3}
    \end{subfigure}
    \hfill
    \begin{subfigure}[t]{0.16\textwidth}
         \centering
         \includegraphics[width=\textwidth]{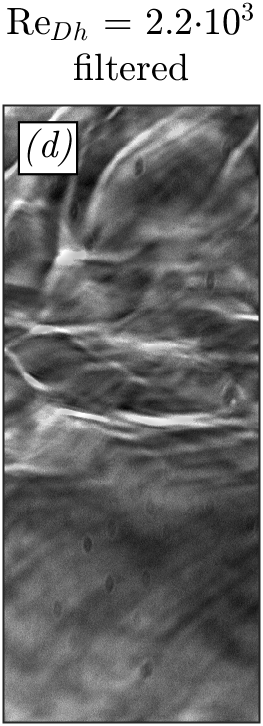}
         %\label{fig: filt4}
    \end{subfigure}
    \hfill
    \begin{subfigure}[t]{0.16\textwidth}
         \centering
         \includegraphics[width=\textwidth]{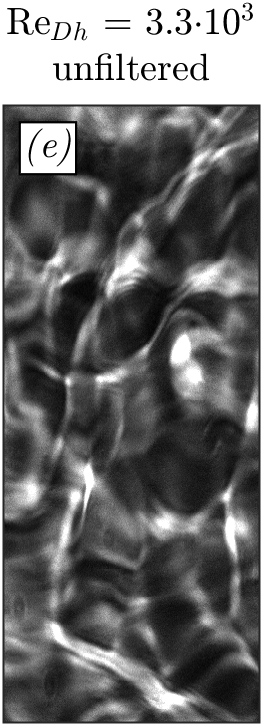}
         %\label{fig: filt5}
    \end{subfigure}
    \hfill
    \begin{subfigure}[t]{0.16\textwidth}
         \centering
         \includegraphics[width=\textwidth]{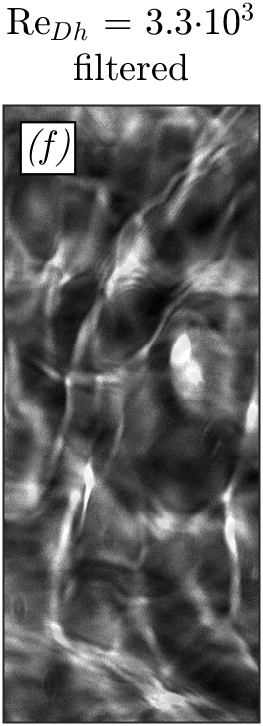}
         %\label{fig: filt6}
    \end{subfigure}
    
\caption{Assessment of effectiveness of current optical filter. The filter removes the background patterns cast by surface finish imperfections of the visors. Optical filter image pairs are shown for case $R1$ in (a) $\&$ (b), case $R3$ in (c) $\&$ (d), and for case $R5$ in (e) $\&$ (f).}
\label{fig:filter}
\end{figure}

The present optical configuration is sensitive to surface imperfections of the visors shown in cross section~\textbf{B} of figure~\ref{fig:schematic}. The imperfections cast optical distortions on the shadowgrams. In order to remove the visual noise induced by the surface flaws and to make the actual compressible structures within the channel discernible in the presented figures, a filter is used. As the large variations in refractive index deform the shadowgraphs and hence the striations cast by the imperfections, subtracting the averaged image from an instantaneous one yields noisy images. Nevertheless, the visors cast striations at several near-constant angles. As such, a spatial filter is tuned to remove a narrow bandwidth of spatial frequencies for a limited range of angles. As such, the filter removes the unwanted striations that can be seen in an averaged shadowgram, whereas structures of other orientations remain unaffected.

The effectiveness of the present filter is assessed in figure~\ref{fig:filter}. The figure compares the unfiltered to the filtered optical signal for a range of Reynolds numbers. At the expense of image contrast, the filter removes most imperfections, whilst keeping the shadows caused by compressible structures mostly intact. The filter is applied to the shadowgrams shown in figures \ref{fig:casebot}, \ref{fig:casetop}, \ref{fig: SSnoheat}a, \ref{fig: SSnoheatv}a, \ref{fig: SSbotheat}a, \ref{fig:plumes}, and \ref{fig: SStopheat}a. However, the filter is not applied to the raw shadowgrams used for the current image velocimetry.
\bibliographystyle{jfm}
\bibliography{jfm}

\end{document}